\documentclass{article}
\usepackage[utf8]{inputenc}
\usepackage{setspace}
\usepackage{lineno}
\usepackage{graphicx}
\usepackage{amsmath,amssymb}
\usepackage{makecell}
\usepackage{multirow}
\usepackage{appendix}
\usepackage{adjustbox}
\usepackage{nameref,hyperref}
\usepackage{multirow}
\graphicspath{{/figures}}
\usepackage{float}
\usepackage{booktabs}
\usepackage[dvipsnames]{xcolor}
\usepackage[square]{natbib}
\usepackage[final]{changes}
\title{Role of Variable Renewable Energy Penetration on Electricity Price and its Volatility Across Independent System Operators in the United States}
\author{Olukunle O. Owolabi$^{1,\dagger}$,
Toryn L. J. Schafer$^{2,\dagger}$, Georgia E. Smits$^{2}$,\\
Sanhita Sengupta$^{3}$, Sean E. Ryan$^{2}$, Lan Wang$^{4}$, David S. Matteson$^{2}$,\\
Mila Getmansky Sherman$^{5\ast}$, Deborah A. Sunter$^{1,6\ast}$ \\
\\
\normalsize{$^{1}$Department of Mechanical Engineering, Tufts University, USA}\\
\normalsize{$^{2}$Department of Statistics and Data Science, Cornell University, USA}\\
\normalsize{$^{3}$School of Statistics, University of Minnesota, USA}\\
\normalsize{$^{4}$Department of Management Science, University of Miami, USA}\\
\normalsize{$^{5}$Isenberg School of Management, UMASS Amherst, USA}\\
\normalsize{$^{6}$Department of Civil and Environmental Engineering, Tufts University, USA}\\
\normalsize{$^{\dagger}$These authors contributed equally to this work.}\\
\\
\normalsize{$^\ast$To whom correspondence should be addressed;} \\ \normalsize{E-mail: Deborah.Sunter@tufts.edu or msherman@isenberg.umass.edu}
}

\date{}

\begin{document}

\maketitle

\section*{Abstract}
The U.S. electrical grid has undergone substantial transformation with increased penetration of wind and solar - forms of variable renewable energy (VRE). Despite the benefits of VRE for decarbonization, it has garnered some controversy for inducing unwanted effects in regional electricity markets. In this study, the role of VRE penetration is examined on the system electricity price and price volatility based on hourly, real-time, historical data from six Independent System Operators (ISOs) in the U.S. using quantile and skew t-distribution regressions. After correcting for temporal effects, \added{we found an increase in VRE penetration is associated with decrease in system electricity price in all ISOs studied.  The increase in VRE penetration is associated with decrease in temporal price volatility in five out of six ISOs studied. The relationships are non-linear}.
\deleted{we found a decrease in price a decrease in price is observed, with non-linear effects on price volatility, for an increase in VRE penetration and non-linear effects of VRE penetration on price volatility.} These results are consistent with the modern portfolio theory where diverse volatile assets may lead to more stable and less risky portfolios.
\\

\textbf{Key Words:} electricity price, electricity price volatility, variable renewable energy, skew-t regression, quantile regression

\section{Introduction}

The U.S. electrical grid is undergoing a transformation with respect to the diversity of assets in its energy portfolio with substantial integration of renewable energy technologies, particularly \replaced{variable renewable energy (VRE)}{VRE} technologies \textemdash non-dispatchable technologies with changing output dependent on renewable resource availability, such as solar and wind power. Although the majority of energy generation comes from fossil fuels, total energy generated from renewable resources (hydroelectric power included) accounted for 18\% of total generation in the United States in 2020 with variable sources - wind and solar energy - accounting for 10.7\% of the total generation  \citep{U.S.EnergyInformationAdministrationEIA2020}. This is owing to increased build-out of renewable energy projects facilitated by the ebbing cost of installation and favorable policy initiatives \citep{U.S.EnergyInformationAdministration2020}. By 2021, an estimated 27.6 GW of renewables is planned to come online with 12.2 GW from wind and 15.4 GW from utility-scale solar \citep{EIA202101}. While the addition of renewable energy will help accelerate the decarbonization of the power sector, its %full-scale realization 
optimal integration will require substantial changes in power system design and market regulations to accommodate the inherent variability of these resources on the electric grid \citep{EIA202101}. To do this, it is important %to understand the effects of high penetration of these 
to elucidate an understanding of the role of VRE on electricity price and price volatility (both with respect to time and VRE penetration), as these impact %and how this affects 
the entire electricity value chain from the energy producers to the consumers.

Several papers have explored the relationship between electricity price and VRE penetration  of wind \citep{Woo2011,Woo2013,BrancucciMartinez-Anido2016,Zarnikau2019,Gil2013,Tsai2018,Quint2019,Haratyk2017,Zarnikau2020}, solar \citep{Deetjen2016,Craig2018}, or both \citep{Bushnell2018,Wiser2017,Barbose2016,Woo2016,Woo2014}. Some of these papers have used simulated data \citep{Haratyk2017,Deetjen2016,Barbose2016}, while others have used historical data \citep{Zarnikau2020,Bushnell2018,Quint2019,Tsai2018,Craig2018,Gil2013,Zarnikau2019,Woo2011,Woo2013,Woo2014,Woo2016}. A few have combined simulated data with historical data validation \citep{BrancucciMartinez-Anido2016,Wiser2017,Mills2020,Mills2021}. The methods used in these studies have primarily been multivariate linear regressions \citep{Zarnikau2020,Bushnell2018,Haratyk2017,Wiser2017,Quint2019,Tsai2018,Craig2018,Zarnikau2019,Woo2011,Woo2013,Woo2014,Woo2016,Mwampashi2020,Gelabert2011} or visualizations with descriptive statistics \citep{Blazquez2018,Cutler2011}. 
It is worth noting that the optimization methods in production cost models for most of the scenario-type/simulation-based analyses \citep{Barbose2016,BrancucciMartinez-Anido2016,Deetjen2016,Seel2018} are often based on linear assumptions. These methods are, however, limited in that they assume a linear relationship between electricity price and VRE with constant variance, but there are inherent distributional changes in the electricity price with respect to VRE requiring more robust modeling choices. One alternative approach has been applied to a case study of PJM\added{(see Table \ref{tab:datasummary} for acronym definitions)} using the Robust Linear Weighted Regression \citep{Gil2013}; however, the scope is limited to a single \replaced{Independent System Operator (ISO)}{ISO} and one year of data.  In general, increasing penetration of VRE is associated with a lower average wholesale price of electricity. This behaviour has been found in both domestic \citep{Zarnikau2020,Bushnell2018,Haratyk2017,Wiser2017,Quint2019,Tsai2018,Craig2018,Barbose2016,Gil2013,Zarnikau2020,BrancucciMartinez-Anido2016,Deetjen2016,Woo2011,Woo2013,Woo2014,Woo2016,Seel2018} and some international \citep{Blazquez2018,Cutler2011,Mwampashi2020,Gelabert2011,Jonsson2010,SaenzdeMiera2008} electricity markets. % since the marginal cost of renewable energy is usually very low.
A summary of the domestic studies can be found in Table \ref{tab:literature}. Using the robust methods in this study, the results not only corroborate the existing literature on increased VRE penetration leading to a reduction in electricity price, but also illustrate that this effect is nonlinear and the greatest impact can be seen in the reductions in extremely high system electricity prices with increased VRE.

Negative system electricity price is also a characteristic of the modern electricity market. Negative prices occur as a result of generation-demand imbalance resulting in high supply during times of low demand. Several factors contribute including a substantial decrease in demand, limited flexibility in power plant operations (e.g., slow or expensive ramping, limited energy storage), and limited transmission capacities.   Studies of some electricity markets have attributed higher frequency of negative prices to high penetrations of VRE \citep{Cutler2011,DeVos2015}. One study found that extreme negative prices - the result of electricity oversupply - are highly correlated with periods of high wind penetration \citep{Cutler2011}. However, there was no relationship observed between the frequency of negative electricity prices and VRE, with inconsistent behaviour observed across the ISOs studied.

While the literature on the average behaviour of electricity price and VRE penetration has been consistent, its price volatility \added{with respect to time (temporal price volatility)} effects \deleted{with respect to time (temporal price volatility)} have been a topic of controversy. Some studies find evidence of an increase in temporal price volatility as VRE penetration increases \citep{Mwampashi2020,Astaneh2013,Woo2011,Seel2018}, while others are unable to show significant evidence that increasing the share of VRE leads to high temporal price volatility \citep{Rai2020,Mulder2013}. A case study of Germany and Denmark found that increasing penetration of VRE could either increase or decrease the temporal price volatility \citep{Rintamaki2017}. While studies have often found that high wind energy penetration contributes to high temporal price volatility \citep{Woo2011,BrancucciMartinez-Anido2016,Mwampashi2020}, a study in South Korea showed that temporal price volatility decreases as wind penetration increases up to 10\% since the wind profile matches the demand patterns at this penetration \citep{Shcherbakova2014}. There are also instances where solar energy penetration has abated temporal price volatility \citep{PereiradaSilva2019,Rintamaki2017}. The evidence from literature suggests that the relationship between temporal price volatility and VRE penetration can vary widely as a result of a confluence of several factors, including patterns of demand \citep{Shcherbakova2014}, weather \citep{Mwampashi2020}, and the availability of flexible generation \citep{Rintamaki2017}. In 5 of the 6 ISOs studied, the system temporal price volatility decreased as the penetration of VRE increased across all quantiles. The results are consistent with the modern portfolio theory that posits that the portfolio of diverse and uncorrelated (or low correlated) assets leads to price volatility reduction\citep{Markowitz1952}.  Similar to the finance field, \added{we show} that adding diverse and volatile assets can lead to a less risky portfolio of such energy assets.  \added{For example, adding solar and wind power to more traditional power generation sources such as coal, natural gas, and nuclear, leads to a more diverse portfolio.  Adding different technologies to the energy portfolio is beneficial, and as a result, an increase in VRE penetration leads to a reduction of temporal price volatility.}

\begin{figure}[t]
\centering
\includegraphics[width=14cm]{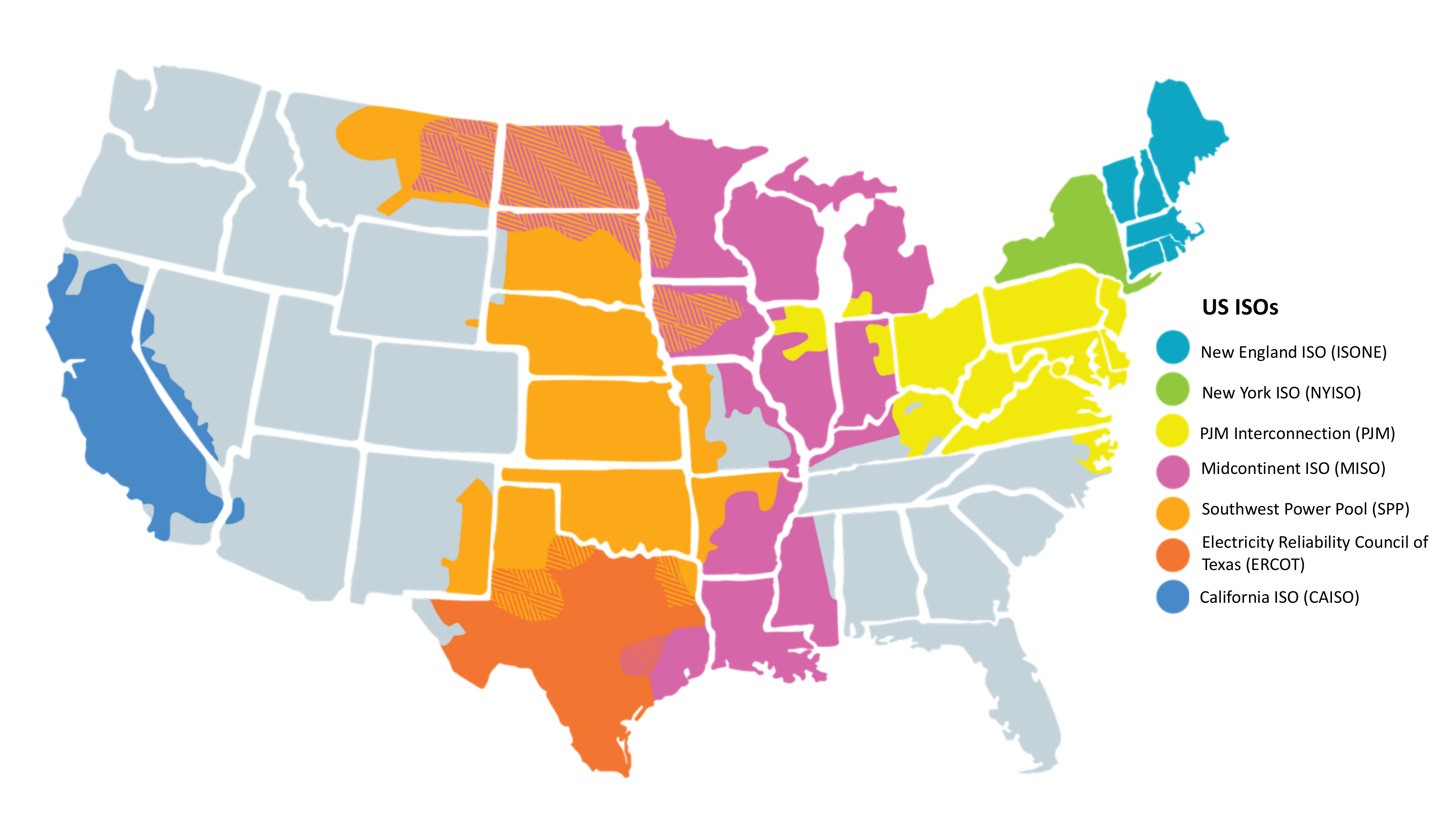}
\caption{Map of Independent System Operators (ISOs) in the United States. ISOs are responsible for coordinating, controlling, and monitoring electricity grid within their region of control. There are 7 ISOs in the United States and the regions covered by these ISOs account for 67\% of the total electricity demand in the country. Adapted from \citet{Climate-XChange2021}.} 
\label{fig:ISO}
\end{figure}

In this study, robust methods - quantile regression and skew-t distribution regression - are used to evaluate the impact of VRE penetration on the conditional electricity price and price volatility. \added{The analysis is done conditional on the seasonality in order to assess effects on the residual price, i.e., we analyze the behavior of the price after removing what is expected given the time of day and year.} Volatility is defined both with respect to time (temporal price volatility) and VRE penetration. Quantile regression and skew-t distribution regression have been chosen since they are well suited for handling outliers, non-linearity, and skewness. \added{By quantifying the skewness of the conditional price distribution, we aim to evaluate the frequency of above and below average conditional prices at varying levels of VRE.}
% (right-skewed with median less than the mean as shown in the Descriptive Statistics - Table 2).
This analysis examines 6 of the 7 \replaced{ISO}{Independent System Operator (ISO)} regions in the United States shown in Figure \ref{fig:ISO} (excluding the Electricity Reliability Council of Texas, ERCOT). At this time, the ERCOT region has been excluded from this analysis due to data constraints that make a comparable analysis infeasible. Despite this, to the best of our knowledge this is the most comprehensive study of the role of VRE on system electricity price and price volatility at the national scale for the United States using robust, nonlinear methods.

%The choice of the United States (with a focus on the Independent System Organizations) as a case study is neither to small (inadequate) nor too large (unreasonable) in such a way to allow a reasonable comparative analysis.
\replaced{Our robust, nonlinear methods for exploring the relationship between VRE and system electricity price and price volatility in several ISOs in the United States could be applied in other markets and has applicability in the broader international context.}{The chosen methodology for exploring the relationship between VRE and system electricity price and price volatility in several ISOs in the United States has applicability in the broader international context.} \replaced{For example, the following}{It is important to note that some} international studies have explored this relationship in Italy \citep{Blazquez2018}, Germany \citep{Blazquez2018, Mwampashi2020}, South Australia \citep{Cutler2011}, Spain \citep{Blazquez2018, SaenzdeMiera2008}, Denmark \citep{Mwampashi2020, Jonsson2010}, and South Korea \citep{Shcherbakova2014} \added{and could be enriched by our chosen methods}. The approach captures non-linearity and distributional effects and can be applied to other electricity markets. \added{Not only the methods, but also} the findings \deleted{also} have a broader context in relation to modern portfolio theory with respect to the addition of a highly volatile asset, VRE, and a corresponding reduction in electricity price and volatility. However, it is important to note that though our methodological approach can be extended to other electricity markets to explore the relationship between VRE and system electricity price or price volatility, the actual trend and relationship may inherently differ in each context owing to several factors, such as the amount of VRE penetration, and the policies and regulations that govern the local electricity market.

\added{This paper aims to \textit{(i)} apply robust, nonlinear methods to the study of electricity price and price volatility, \textit{(ii)} elucidate more generalizable conclusions by scaling the analysis to explore multiple ISOs, and \textit{(iii)} contextualize the results based on modern portfolio theory.}

\begin{table}[!htp]
% \renewcommand{\thetable}{S\arabic{table}}
% \centering
\begin{adjustbox}{width=1\textwidth}
\begin{tabular}{| c | c | c | c |}
% \toprule
\hline
\textbf{Author} & \textbf{Region} & \textbf{Time Period} & \makecell{ \\\textbf{Decrease in System Electricity Price}\\\textbf{per GWh of \added{Specified} VRE}}\\
\hline
 \citet{Zarnikau2020} & MISO & \makecell{2013-2017} & \makecell{Wind (DAM): \$2.1-\$6}\\
\hline
\citet{Bushnell2018} & CAISO & \makecell{2012-2016} & Solar: \$0.1\\
\hline 
\citet{Haratyk2017}& \makecell{Midwest, \\Mid Atlantic} & \makecell{2008-2015} & \makecell{Wind (midwest): \$0.612, \\Wind (Mid atlantic): \$ 0}\\
\hline

 \citet{Quint2019}& MISO & \makecell{2008-2016} & \makecell{Wind: \$1.4 - \$3.4} \\
  \hline
\citet{Tsai2018} & ERCOT & \makecell{2014-2016} & \makecell{Wind: \$1.45-\$4.45}\\

 \hline
 \citet{Zarnikau2019}& ERCOT & \makecell{2011-2017} & Wind: \$1.64\\
\hline

  \citet{Woo2011}& ERCOT & \makecell{2007-2010} & \makecell{Wind (Houston):\$3.9\\
Wind (North):\$6.1 \\ Wind (South):\$3.2 \\ Wind (West):\$15.3} \\
\hline
 \citet{Woo2013}& Mid-C Hub & \makecell{2006-2012}  & \makecell{Wind (Day):\$0.96\\
Wind (Night):\$0.72 }\\
\hline
 \citet{Woo2014}& CAISO & \makecell{2010-2012} & \makecell{Solar (NP15):\$12.4\\
Wind (NP15):\$7.8 \\ Solar (SP15):\$12.2 \\ Wind (SP15):\$9.8 \\ Solar (ZP26):\$14.3 \\ Wind (ZP26):\$7.9 } \\
\hline
 \citet{Woo2016}& CAISO & \makecell{2012-2015} & \makecell{RTM\\
Solar (NP15):\$2.2\\
Wind (NP15):\$2.8 \\ Solar (SP15):\$3.7 \\ Wind (SP15):\$1.5\\
DAM\\
Solar (NP15):\$5.3\\
Wind (NP15):\$3.3 \\ Solar (SP15):\$3.2 \\ Wind (SP15):\$1.4}  \\
\hline
\end{tabular}
\end{adjustbox}
    \caption{Literature summary for US studies (domestic) on VRE impact on price. \added{The specified VRE, solar or wind, is given in the rightmost column.} (note: RTM - real time market, DAM - day ahead market)}
    \label{tab:literature}
\end{table}

\section{Methods}

\subsection{Data Collection \& Processing}
% \nocite{*[resetnumbers=1]}
Each Independent System Operator (ISO) collects data on the system price of electricity and the amount of electricity that was generated from various energy sources (e.g., solar, wind, natural gas, coal, nuclear).
These data are available on each ISO's respective website in various formats.
For example, some ISOs release files containing full years worth of data,
while others provide daily updates through widgets.
Therefore, a web scraper was built to collect the data for each ISO.
Note the different operators collect data at different temporal frequencies,
but the highest resolution %smallest time unit 
that the operator provides is always used. Furthermore, the time periods covered for each operator do not match. 
As a result, there are more data for some operators than others;
however, this difference is not expected to impact the analysis \added{as each analysis is done independently} (See Table \ref{tab:datasummary} for the temporal resolution and coverage period for each ISO). From this price and supply information, we calculate two variables at each time point:
\emph{(i)} the percentage of energy generated using VRE
and \emph{(ii)}  the system price per megawatt-hour of electricity.

\begin{table}[htp]
\centering
\begin{adjustbox}{width=1\textwidth}
\begin{tabular}{ |c | c | c | c | c | }
% \toprule
\hline
     \textbf{ISO} & \textbf{date range} & \makecell{\textbf{original}\\ \textbf{resolution}}& \textbf{VRE} & \textbf{Reference}   \\
     \hline
% \midrule
   \makecell{{New England Independent}\\{ System Operator, ISONE}}  & 2014 - 2020 & hourly  & \makecell{{solar}\\ {and wind}} &  \citet{ISONE2021}  \\\hline
    \makecell{{New York Independent}\\ {System Operator, NYISO}}    & 2015 - 2020 &   5 mins & wind  & \citet{NYISO2020}  \\\hline
    \makecell{{Pennsylvania, Jersey, Maryland}\\ {Power Pool, PJM}}   & 2016 - 2018  & hourly      & \makecell{{solar}\\ {and wind}} &\citet{PJM2021}\\\hline
    \makecell{{Midcontinent Independent}\\ {System Operator, MISO}}   & 2015 - 2019     & hourly   &  Wind & \citet{MISO2021} \\\hline
   Southwest Power Pool, SPP   & 2019 - 2020  & 5 mins &   \makecell{{solar}\\ {and wind}}& \citet{Spp2021}    \\\hline
    \makecell{{California Independent}\\ {System Operator, CAISO}}   & 2017 - 2020     & 5 mins   & \makecell{{solar}\\ {and wind}} & \citet{CaliforniaISO2017}  \\\hline
% \bottomrule                          
\end{tabular}
\end{adjustbox}
\caption{Temporal range, resolution, and VRE composition for ISO data}
\label{tab:datasummary}
\end{table}
\subsubsection{Percentage of VRE}
In calculating the percentage of VRE used throughout the analysis, only electricity generated from solar and wind is considered. However, each ISO provides slightly different granularity on the energy sources in their electricity generation mix. For example, if the relative generation from solar is low, then this source may be bundled with other renewable energy sources. Because of the variability of solar adoption and the differences in data reporting among ISOs, solar electricity generation data was not specifically reported in the generation mix for NYISO and MISO, in which case only the wind electricity generation data was used to represent the VRE percentage (Table \ref{tab:datasummary}). Note, the effect of these differences is small and should not impact the analysis. The percentage of VRE is calculated as the sum of the generation from solar and wind sources divided by the total amount of generation from all sources.

\subsubsection{System Price}
%Computing the price per watt of energy is more complicated.
Each ISO reports information on the price paid at specific locations or hubs in their network.
Due to transmission and congestion costs, the price at different hubs can be different.
To ensure that the results are interpretable and comparable the transmission and traffic/congestion costs is removed to compute a system price for each ISO.

For each ISO, this system price reflects the cost of generating one megawatt-hour of electricity.
Again, since each ISO has different reporting standards,
there are slight differences in how the system price is computed for each ISO.
Full details for the relevant procedure for each ISO are given in Appendix \ref{sec:iso}.
As before, these differences \replaced{are}{is} not expected to impact the analysis.

\subsection{Seasonal  and Diurnal Adjustment}
Changes in energy price are driven to a large extent by changes in consumer demand. 
For example, energy prices are higher during the day (when consumers are awake) than at night and higher during the summer (when consumers use air conditioning) than spring.
To properly analyse the impact of VRE penetration on energy prices, 
this \added{expected} pattern is taken into account \added{and used to detrend the price}. \added{Before detrending prices at the temporal scales, it is not clear whether the estimated associations between energy price and VRE penetration are due to expected seasonal fluctuations or changes in VRE penetration. By removing the expected pattern, we can analyze the effect of VRE penetration on the changes in energy price compared to what is expected.} 

\begin{figure}[htp]
\centering
\includegraphics[width=\textwidth]{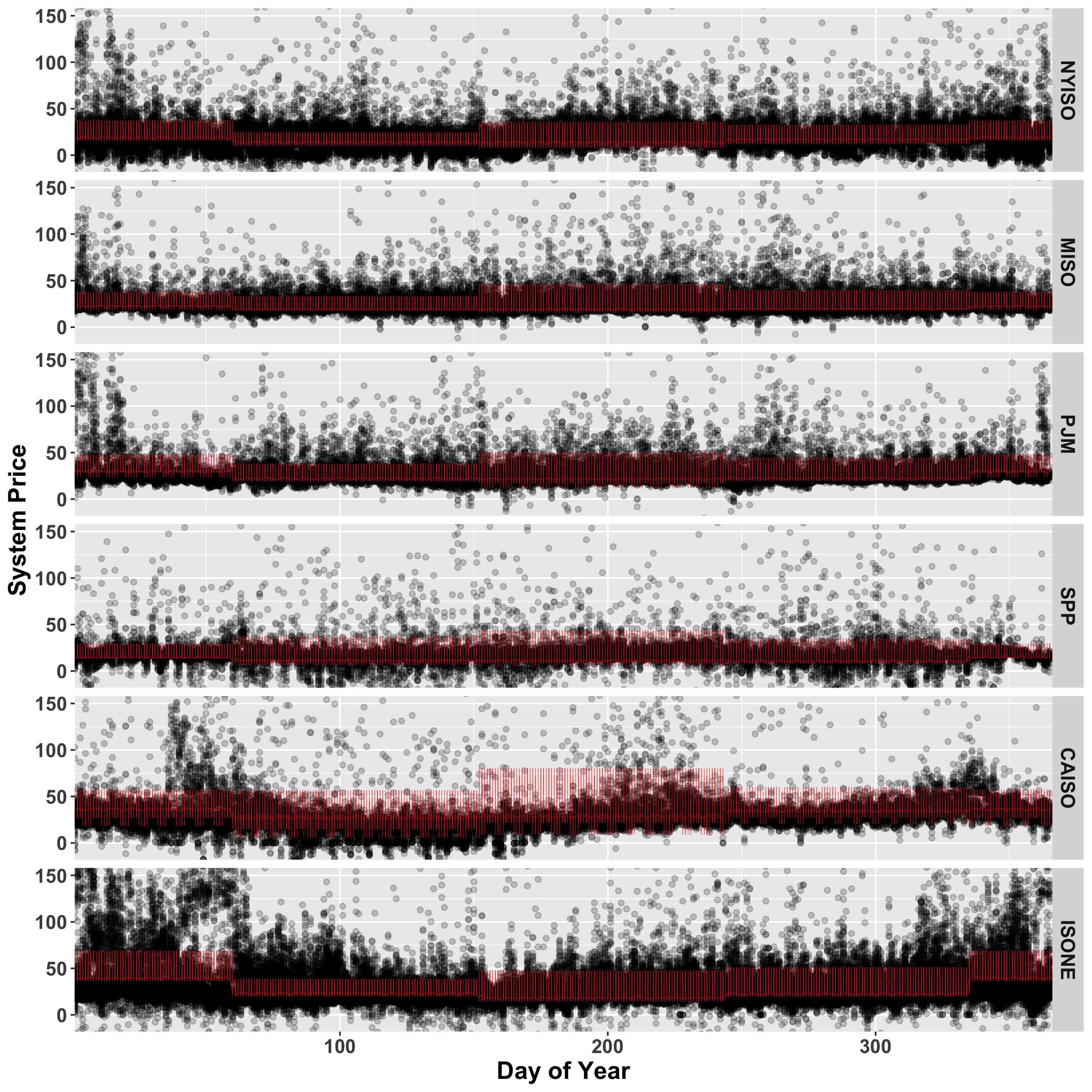} % changed width from 14cm to textwidth so that figure would be centered - GA
\caption{\replaced{The energy price data for each ISO is detrended by the average price for hour, season, and weekend trends. The observed price data, shown with black dots, is plotted by day of the year. The red line indicates the estimated trend. The resulting detrended price is the observed value minus the average. Note there are repeated observations for multiple years and the y-axis is clipped to the range -\$10 to \$150.}{Correction for temporal effects. Shown here in red is the diurnal and seasonal effect on the energy price data for each ISO (in \$). This correction is based on subtracting the average diurnal pattern over defined time periods of the year corresponding to spring, summer, fall and winter seasons.}}  
\label{fig:seasonal_correction}
\end{figure}

\added{We define what is expected as} a linear model with categorical variables \deleted{is} used to estimate these temporal effects. 
For the hourly data, categorical variables for hour of day, season, and weekend are included.
The model allows for interaction effects between the hour of day and season, 
so that each season has its own diurnal pattern. 
More formally, for the hourly data the price at time $t$ is given by  
\begin{equation}\label{eq:hourly_lm}
    \text{Price}(t) = 
\beta_1 \text{hour}(t) + \beta_2 \text{season}(t) + \beta_3 \text{hour}(t)\times \text{season}(t) + \beta_{4} \text{weekend}(t) + \epsilon_t 
\end{equation}
where hour$(t)$, season$(t)$ give the hour of the day and season at time $t$ respectively,
weekend$(t)$ is a binary variable indicating whether time $t$ occurs on a weekend and 
$\epsilon_t$ is a residual term. \added{There are four seasons which correspond to the following calendar month groupings: Winter = {12, 1, 2}, Spring = {3, 4, 5}, Summer = {6, 7, 8}, and Autumn = {9, 10, 11}.}
Note, in a standard linear model the residual term is noise.
This is not the case in this setting
as the price (and temporal price volatility) is expected to be driven by varying levels of VRE.
Fitted values for the 6 locations are shown in Figure \ref{fig:seasonal_correction}.
The estimated diurnal pattern for CAISO (separated by season) is shown in Figure \ref{fig::caiso_diurnal}. A more detailed figure is available in the Appendix \ref{appendix:detrend}.

The parameters were estimated using the base \textit{lm} function within the R programming language.
The residual terms, $\epsilon_t$, are  used to explore
the relationship between VRE and price, and is referred to as the detrended system electricity price. The following analyses will make use of this detrended system price to calculate temporal volatility as well as examine the relationship \replaced{with}{between} VRE penetration.

\subsection{Temporal Price Volatility}
\added{
We calculated the temporal volatility of detrended system electricity price as a function of time (temporal price volatility), using the Exponential Weighted Moving Standard Deviation (EWMSD). 
Detrended price volatility is the instantaneous standard deviation of the hourly detrended electricity price. The exponentially weighted moving standard deviation improves on the sample standard deviation by assigning weights to the periodic price. Specifically, EWMSD gives more weight to the recent price and decreases the weights exponentially as historical data points increases. 
The equation is usually given as:}

\begin{equation}\label{eq:ewmsd}
    \added{EWMSD_{i} = (1-\lambda)U_i^2 + (1-\lambda)\lambda^2 U_{i-1}^2 + ... +  (1-\lambda)\lambda^n U_{i-n}^2}
\end{equation}
\added{where
$U_i^2$ is the price variance in the current hour $i$ and 
$\lambda$ is the weight. We have used a weight of 0.94 since this a standard weighting factor used in financial risk/volatility analysis} \citep{longerstaey1996riskmetricstm}. \added{This method of calculating volatility is based on the operation stage volatility (realised effects) and should not be confused with planning stage volatility that attempts to capture differences between the predicted and observed real-time price variations } \citep[predicted effects;][]{Astaneh2013}.

\subsection{Nonparametric quantile regression}
A nonparametric quantile regression model was fitted based on B-splines \added{basis function expansion of percentage VRE} to examine the possible nonlinear effects of the percentage of VRE on the hourly electricity price and temporal price volatility. This approach does not require us to assume a specified distribution \added{for the response} and is robust with \deleted{respect to} heavy-tailed distributions \added{as compared to linear regression}. By fitting quantile regression at different quantile levels, a more comprehensive characterization of \added{the} effect of the percentage of renewables \added{on price and price volatility} can be obtained.  

\added{The $\tau$th ($0<\tau<1$) quantile of a random variable $Y$ is $Q_{Y}(\tau) = \inf\{t:F_{Y}(t)\geq\tau\}$ where $F_Y$ is the cumulative distribution function of $Y$. Quantiles in the context of regression had been first introduced in \citet{koenker1978}. Given a dependent variable $Y$ and independent variable matrix $X$, linear regression estimates the conditional mean of $Y$ given $X$ by assuming the dependence relation $Y = X'\beta$ and that the cumulative distribution of $Y$ given $X$, $F_{Y|x}$, is a Gaussian distribution among other assumptions. Instead, quantile regression assumes that the $\tau$th ($0<\tau<1$) conditional quantile of $Y$ given $X=x$ is defined as $Q_{Y}(\tau|x) = \inf\{t:F_{Y|x}(t)\geq\tau\} = X' \beta$ without any further assumption on $F_{Y|x}$. The conditional median corresponds to $Q_{Y}(0.5|x)$.  Interpretation of the conditional quantile is straightforward.  For example, given the predictor value $X=x$ and $\tau=0.9$, 90\% of observations of $Y$ with associated $x$ fall below $Q_{Y|X}(0.9)$. Given $\tau=0.9$, an increase in $X$ by 1 unit would increase the 0.9 quantile of $Y$ by $\beta$.}

\added{In general, assuming the linear functional form $X'\beta$ of the quantile $Q_{Y}(\tau|x)$ may lead to undesirable results when the conditional quantile of $Y$ given $X$, $Q_{Y}(\tau|x)$, is not linear in $X$ and hence misspecified. Splines can be used as a curve fitting technique for $Q_{Y}(\tau|x)$ using piecewise polynomials of $X$. Knots are values of the independent variable which are used to partition the curve into pieces to fit a final piecewise polynomial and the degree of each polynomial defines the degree of the spline. The nonparametric quantile regression models the conditional quantile using a spline approximation instead of a specific functional form of $X$. The B-splines \citep{Schoenberg1988, Curry1988} approximation is known to be flexible and computationally efficient. Specifically, let $\pi(t)= \left(b_1(t),...,b_{k_n+l+1}(t)\right)'$ denote a vector of normalized B-spline basis functions of order $l+1$ and degree $l$. Then $Q_{Y}(\tau|x)$ is approximated by $\beta_0 + \pi(x)'\beta$, where $\beta_0$ is the intercept and $\beta$ are a vector of $l+1$ coefficients to be estimated from the data by minimizing a quantile loss function. The quantile regression is implemented using the \textit{quantreg} package in R which can incorporate splines. Splines can be calculated using the \textit{splines} package in R.}

\added{When analyzing the hourly price, the nonparametric quantile regression is applied to the detrended hourly price (dependent variable) conditional on B-splines of percentage of VRE (independent variables), as previously described. When analyzing the hourly temporal price volatility, nonparametric quantile regression is applied to the exponentially weighted moving standard deviation (EWMSD) estimator of hourly volatility \eqref{eq:ewmsd}(dependent variable), computed using the detrended price \eqref{eq:hourly_lm} with B-splines of percentage of VRE as the independent variables. B-splines of degree 3 is used with number of knots depending on visual inspection of the data scatterplot of price or price volatility vs VRE such that the splines do not overfit at the boundaries.}

\added{We evaluate the subsequent non-linear effects induced by the B-spline basis by calculating an approximate derivative. The approximate derivative of the spline basis functions, $\pi'(t)= \left(b_1'(t),...,b_{k_n+l+1}'(t)\right)'$ can be calculated using the \textit{JM} package in R. The derivative of $Q_{Y}(\tau|x)$ is then approximated by $(\pi'(x))'\beta$.} \\

\deleted{Given a response variable $Y$ and a predictor $X$, quantile regression estimates the effects of $X$ on the conditional quantile of $Y$ \added{given $X$}.  Formally, the $\tau$th ($0<\tau<1$) conditional quantile of $Y$ given $X=x$ is defined as $Q_{Y}(\tau|x) = \inf\{t:F_{Y|x}(t)\geq\tau\}$, where $F_{Y|x}$ is the conditional cumulative distribution function of $Y$ given $X=x$. The conditional median corresponds to $Q_{Y|x}(0.5)$.  Interpretation of the conditional quantile is straightforward.  For example, given the predictor value $X=x$ and $\tau=0.9$, 90\% of observations of $Y$ with associated $x$ fall below $Q_{Y|X}(0.9)$. }

\deleted{The nonparametric quantile regression models the conditional quantile \replaced{non-linearly}{nolinearly} using a spline approximation. Specifically, let $\pi(t)= \left(b_1(t),...,b_{k_n+l+1}(t)\right)'$ denote a vector of normalized B-spline basis functions of order $l+1$. Then $Q_{Y}(\tau|x)$ is approximated by $\pi(x)'\beta$, where $\beta$ are \added{a vector of $l+1$ coefficients} to be estimated from the data by minimizing a quantile loss function. The B-spline approximation is known to be flexible and computationally efficient. }

\deleted{When analyzing the hourly price, the nonparametric quantile regression is applied to the detrended hourly price, as \added{previously }described\deleted{in the previous section}. When analyzing the hourly temporal price volatility, nonparametric quantile regression is applied to the exponentially weighted moving standard deviation (EWMSD) estimator of hourly volatility \eqref{eq:ewmsd}, computed using the detrended price \eqref{eq:hourly_lm}. } \deleted{Volatility is the instantaneous standard deviation of electricity price. The exponentially weighted moving average improves on simple variance by assigning weights to the periodic price. Specially, EWMA gives more weight to the recent price and decreases the weights exponentially as the data points become further away.}

\noindent\textbf{Skew t-distribution regression} \\

\added{It was clear from plots of price against VRE penetration (Figure \ref{fig:sst_quant}) that many features of the distribution of price were changing as a function of VRE penetration. As an alternative to the nonparametric quantile regression, we chose to additionally model the distribution with the parameteric skew t-distribution. The skew t-distribution analysis provides complementary interpretation of generalized linear effects on the conditional distribution of detrended system price given VRE penetration.} 

A generalized linear model was fitted for location, shape, and scale \replaced{of}{for} the conditional distribution of detrended price, $\epsilon_t$, given percentage VRE. We used the four parameter skew t-distribution  (ST) parameterized as in \citet{fernandez1998bayesian} with parameters $- \infty < \mu < \infty$, $\sigma > 0$, $\nu > 0$, and $\tau > 0$. The centrality and skewness (shape) parameters, $\mu$ and $\nu$ respectively, varied with percent VRE ($x_t$) via a generalized linear model:
    \begin{align*}
     \mu_t &= \beta_0^{\mu} + \beta_1^{\mu} x_t/100, \\
     \log(\nu_t) &= \beta_0^{\nu} + \beta_1^{\nu} x_t/100,
    \end{align*}
where the log transformation ensures positivity of $\nu_t$ and the scale and tail index parameters, $\sigma$ and $\tau$ respectively, were assumed to be constant: 
    \begin{align*}
     \log(\sigma) &= \beta_0^{\sigma}, \\
     \log(\tau) &= \beta_0^{\tau}. 
    \end{align*}
Therefore, the following \added{conditional} distribution \replaced{is}{are} obtained for the residual price: $\epsilon_t \vert x_t \sim \textrm{ST}(\mu_t, \nu_t, \sigma, \tau)$. All parameters were estimated by maximum likelihood with the RS algorithm as implemented in the R package \textit{gamlss} \citep{stasinopoulos2007generalized}. \added{For a given level of VRE, $x_t$, we evaluate the estimated distribution by quantile summaries as the complete distribution is defined.}

\begin{figure}[!h]
    \centering
    \includegraphics[width=14cm]{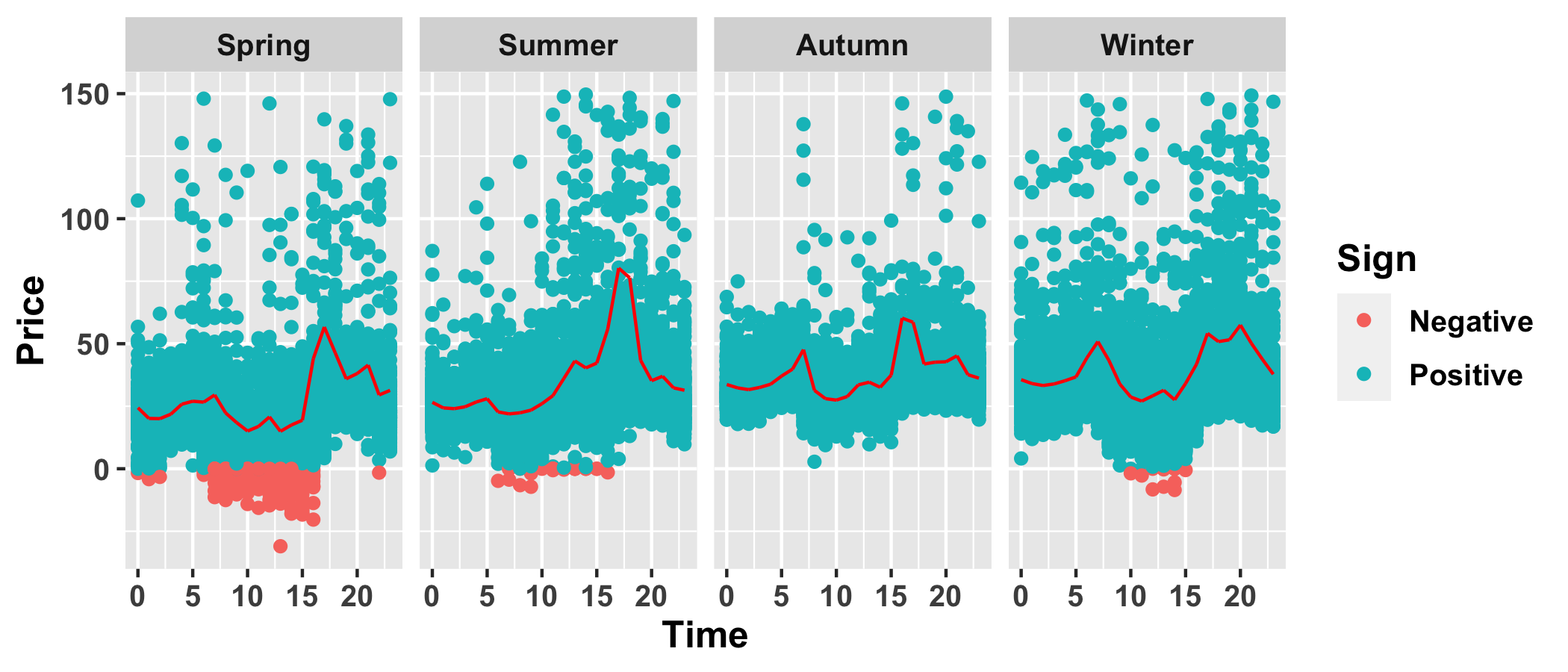} % old: caiso_diurnal.pdf
    \caption{CAISO raw system prices by seasons and hour of day. The plot shows the estimated diurnal pattern (red line) by season. For plotting purposes prices greater than 150 are excluded. Orange dots depict negative prices and teal dots depict positive prices (all in \replaced{USD}{\$}).}
    \label{fig::caiso_diurnal}
\end{figure}

\begin{figure}[h]
\centering
\includegraphics[width=8cm]{ 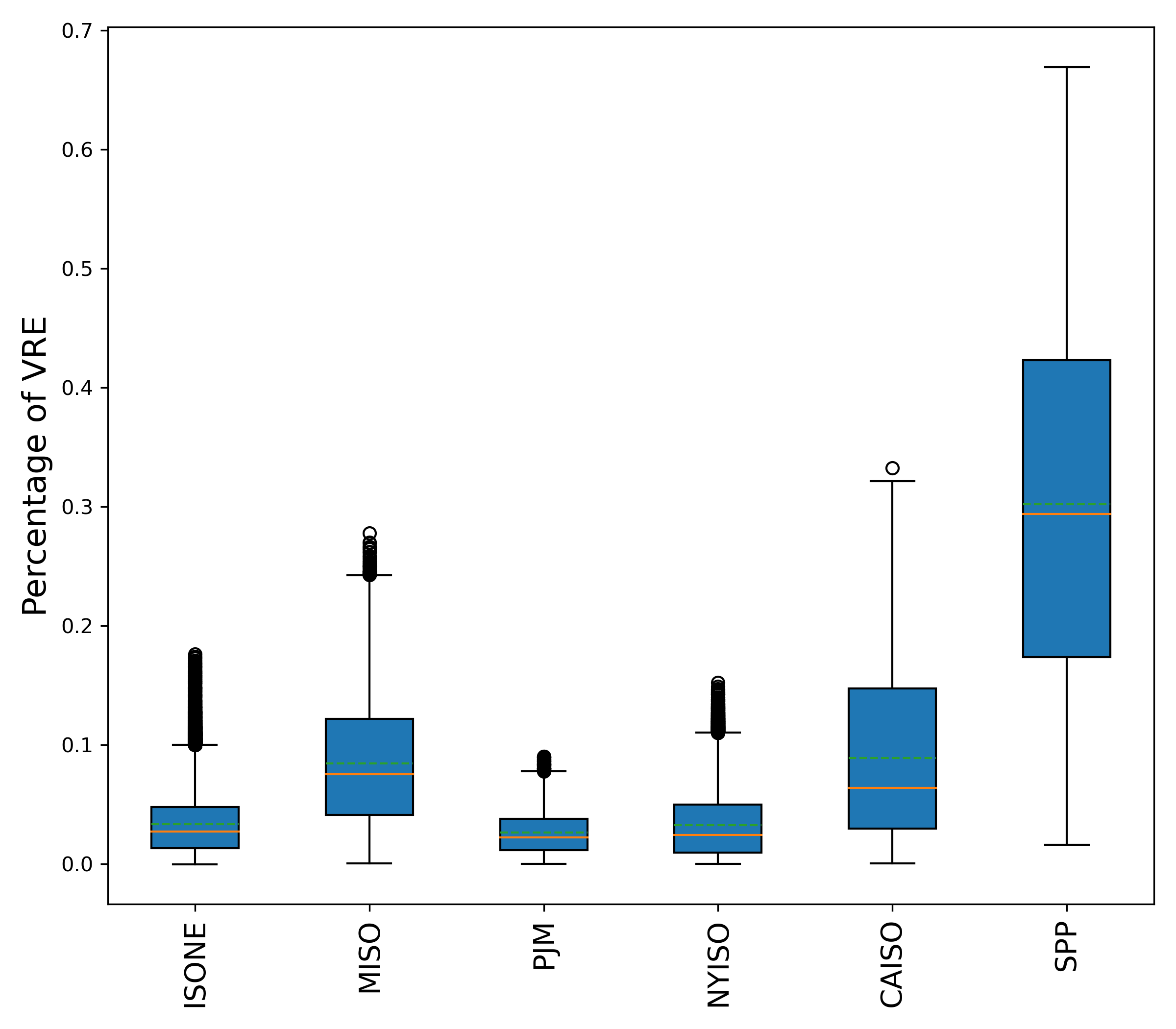}
\caption{VRE (Solar and Wind) penetration. Shown here is the percentage of VRE generation to total generation in each of the 6 ISOs considered. Southwest Power Pool (SPP) achieved the highest VRE penetration based on the data used and for the study period considered, followed by CAISO, MISO, ISONE, NYISO and PJM.} 
\label{fig:VRE}
\end{figure}

 \section{VRE Penetration and System Electricity Price}
Renewable energy resources and the adoption of utility-scale VRE technologies varies substantially throughout the United States, as seen in the corresponding VRE generation shares across ISOs in Figure \ref{fig:VRE}. In order to accurately assess the role of VRE on system electricity price, a comparative analysis across multiple ISOs was performed using regional data from 2014-2020 when available. Details on the regional data used for each ISO can be found in Table \ref{tab:datasummary}. The system electricity price is based on the real-time wholesale price of electricity excluding the congestion costs and transmission losses. Corrections were also made for the diurnal, seasonal, and weekend effects that may influence demand and resource availability and ultimately impact price (shown in Figure \ref{fig::caiso_diurnal}). Descriptive statistics for the electricity price and temporal price volatility after applying these corrections, can be found in Table \ref{desc stat2}. For each ISO studied, the mean system electricity price is substantially larger than the median price, indicating skewness and price outliers, suggesting that the ordinary least squares approach applied in the majority of the previous studies using multivariate linear regression \citep{Zarnikau2020,Bushnell2018,Haratyk2017,Wiser2017,Quint2019,Tsai2018,Craig2018,Zarnikau2019,Woo2011,Woo2013,Woo2014,Woo2016,Mwampashi2020,Gelabert2011} may not have adequately captured the dynamics of the existing relationship between VRE penetration and electricity price.

%\begin{table}[htp]
\begin{table}[h]
\centering
\begin{tabular}{llrrrrrr}
\toprule
 &  & CAISO & ISONE & MISO & NYISO & PJM & SPP\\
\midrule
 & min & -76.29 & -208.37 & -33.51 & -2504.87 & -78.80 & -49.36\\

 & median & -5.14 & -4.82 & -1.74 & -1.02 & -2.79 & -2.63\\

 & max & 957.62 & 2396.98 & 546.73 & 1145.75 & 615.95 & 1086.95\\

 & mean & 0.00 & 0.00 & 0.00 & 0.00 & 0.00 & 0.00\\

\multirow{-5}{*}{\raggedright\arraybackslash Detrended Price (\$)} & st. dev. & 41.34 & 34.26 & 14.73 & 36.26 & 20.28 & 27.27\\
\cmidrule{1-8}
 & min & 0.19 & 0.15 & 0.13 & 0.18 & 0.11 & 0.23\\

 & median & 1.83 & 1.67 & 1.17 & 1.59 & 1.36 & 1.72\\

 & max & 24.92 & 33.79 & 18.69 & 37.21 & 19.30 & 27.50\\

 & mean & 2.36 & 1.95 & 1.49 & 1.93 & 1.72 & 2.17\\

\multirow{-5}{*}{\raggedright\arraybackslash Price Volatility (\$)} & st. dev. & 2.13 & 1.29 & 1.18 & 1.52 & 1.26 & 1.73\\
\cmidrule{1-8}
 & min & 0.04 & 0.00 & 0.04 & 0.00 & 0.01 & 3.10\\

 & median & 7.28 & 2.74 & 7.56 & 2.44 & 2.22 & 48.15\\

 & max & 38.51 & 17.64 & 27.78 & 15.24 & 9.16 & 83.30\\

 & mean & 10.04 & 3.37 & 8.47 & 3.27 & 2.62 & 46.38\\

\multirow{-5}{*}{\raggedright\arraybackslash Percentage of VRE (\%)} & st. dev. & 8.10 & 2.58 & 5.24 & 2.82 & 1.79 & 16.81\\
\bottomrule
\end{tabular}

\caption{Descriptive Statistics Summary. This shows the minimum, median, maximum and mean statistics for the detrended  system price (in \$) and temporal price volatility (in \$). (Note: sd = standard deviation)}
\label{desc stat2}
\end{table}

\begin{figure}[h!]
\centering
\includegraphics[width=14cm]{ 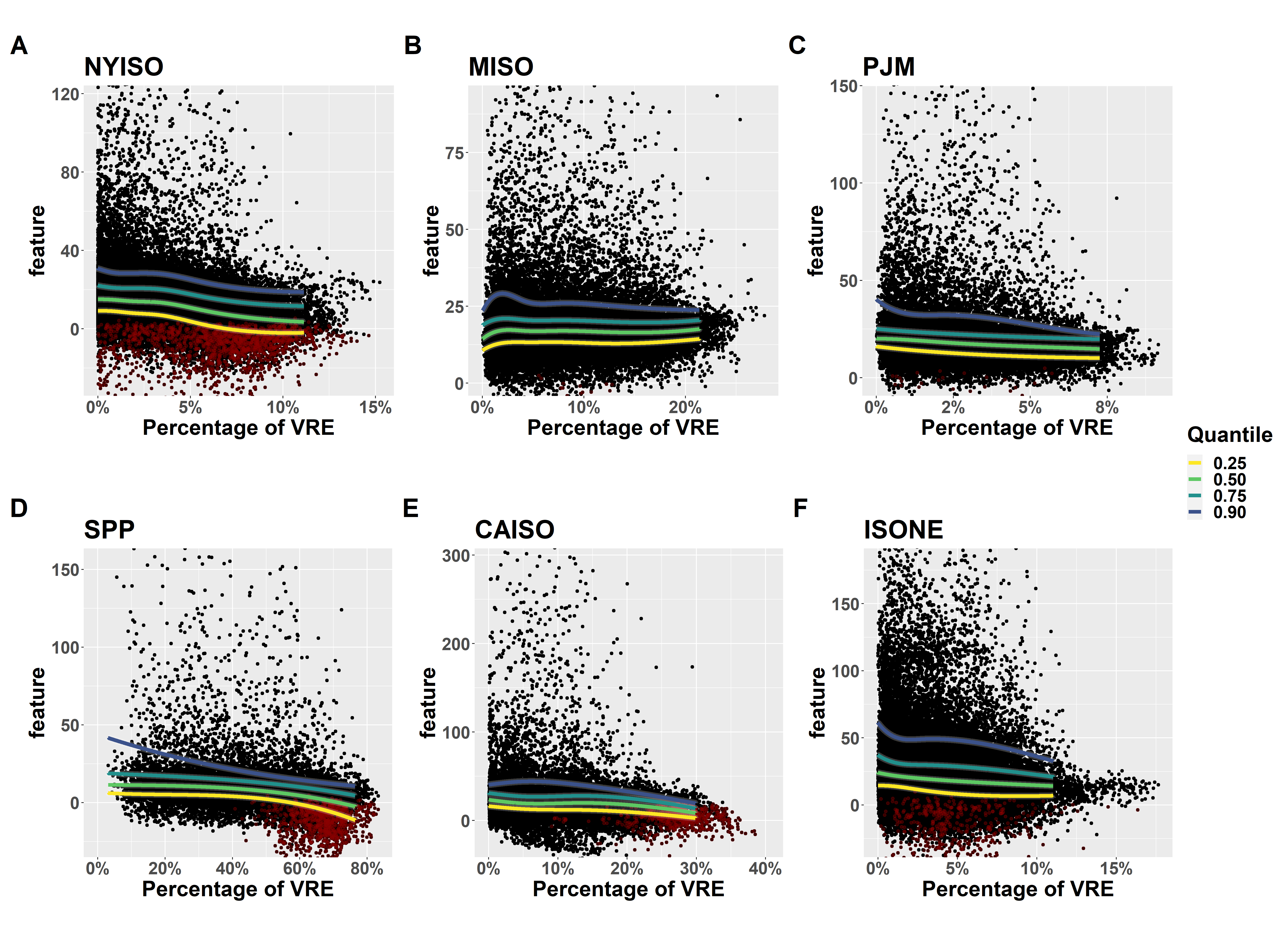}
\caption{VRE penetration (in \%) and Detrended Electricity Price (in \$). \replaced{The six panels}{A, B, C, D, E and F} show the relationship between VRE penetration and detrended electricity price for NYISO, MISO, PJM, CAISO, ISONE and SPP respectively. The continuous lines show the quantiles with the colors - yellow, green, blue and purple representing the \replaced{0.25th, 0.50th (median), 0.75th and the 0.90th quantiles}{25th, 50th (median), 75th and the 90th percentiles} respectively. Each panel's axis limits are adjusted separately in order to clearly show the quantiles' trend direction.} 
\label{fig:quantile_price}
\end{figure}

The quantile regression results in Figure \ref{fig:quantile_price} show that an increase in the penetration of VRE is associated with a  decrease in detrended system electricity price. While this trend is generally true across all ISOs and quantiles \replaced{(0.25th, 0.50th, and 0.75th)}{(25\%, 50\%, and 75\%)}, the relative system price reduction is often greatest in the extreme prices, i.e., the \replaced{0.90th}{90\%} quantile. Additionally, the change in the system price is not constant across the percentage of VRE nor the quantiles and, once again, this is typically more apparent at higher quantiles. For example, in PJM, Figure

\begin{figure}[htp]
\centering
\includegraphics[width=13cm]{ 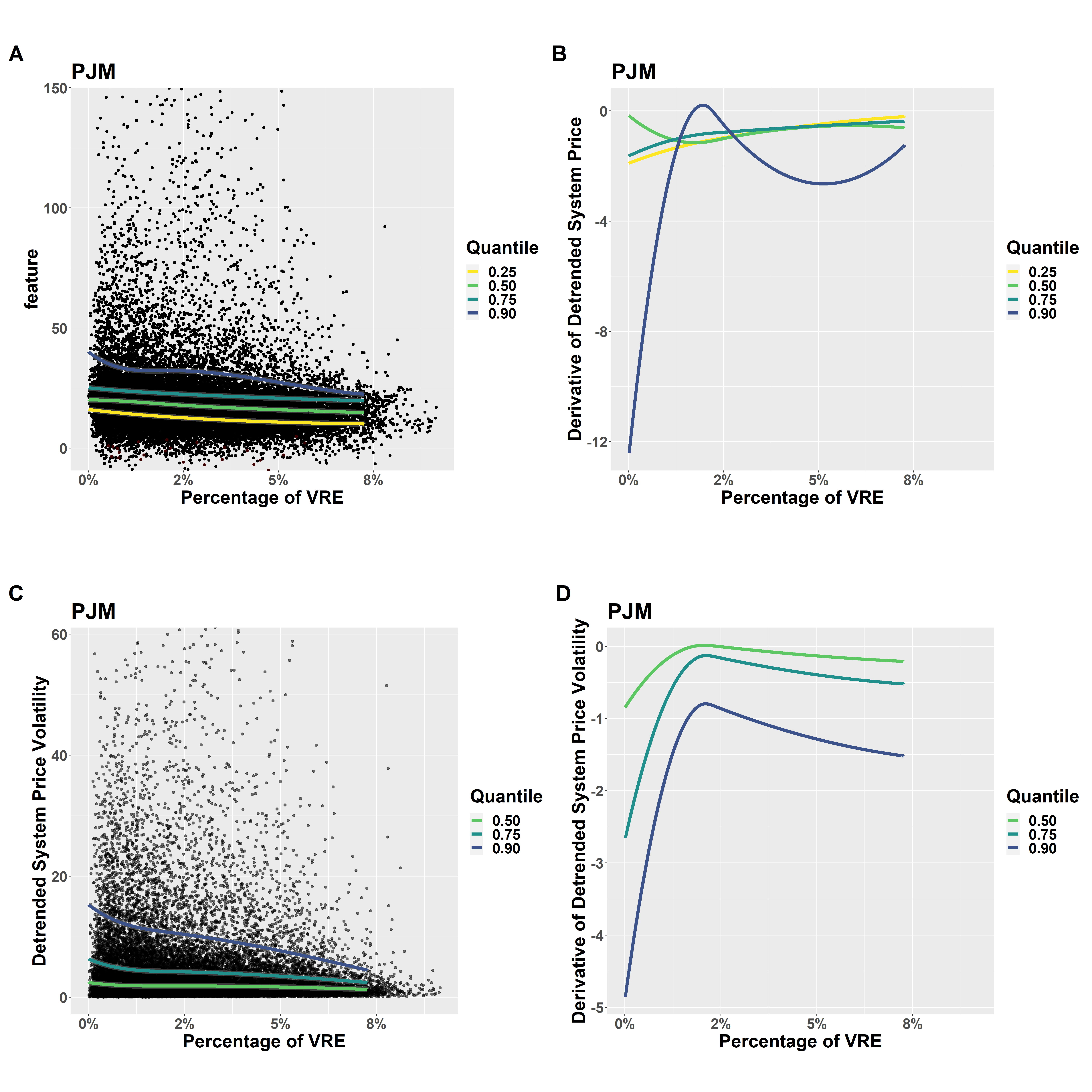}
\caption{The derivative of the detrended system price (in \$) and detrended system price volatility (temporal volatility in \$) with respect to the percentage of VRE for PJM Independent System Operator.} 
\label{fig:derivative}
\end{figure}

\ref{fig:quantile_price}C, when the VRE percentage changes from 0-1\% the median detrended system price decreases by \$1.6 USD in average absolute values and the \replaced{0.90th quantile}{90th percentile} of the detrended system price decrease by \$10 in average absolute values, but when the VRE percentage changes from 5-6\% the median detrended system price decreases by \$0.2 in average absolute values and the \replaced{0.90th quantile}{90th percentile} detrended price decreases by \$0.8 in average absolute values. The derivative of the detrended system price with respect to the percentage of VRE is depicted in Figure \ref{fig:derivative} \added{for PJM}. 
It is evident that increasing penetration of VRE lowers the frequency of occurrence of higher extremes of system prices \added{as \ref{fig:derivative}A-B suggests the value of the \replaced{0.90th quantile}{90th percentile} is decreasing with increasing VRE}.

\begin{figure}[h!]
\centering
\includegraphics[width=13cm]{ 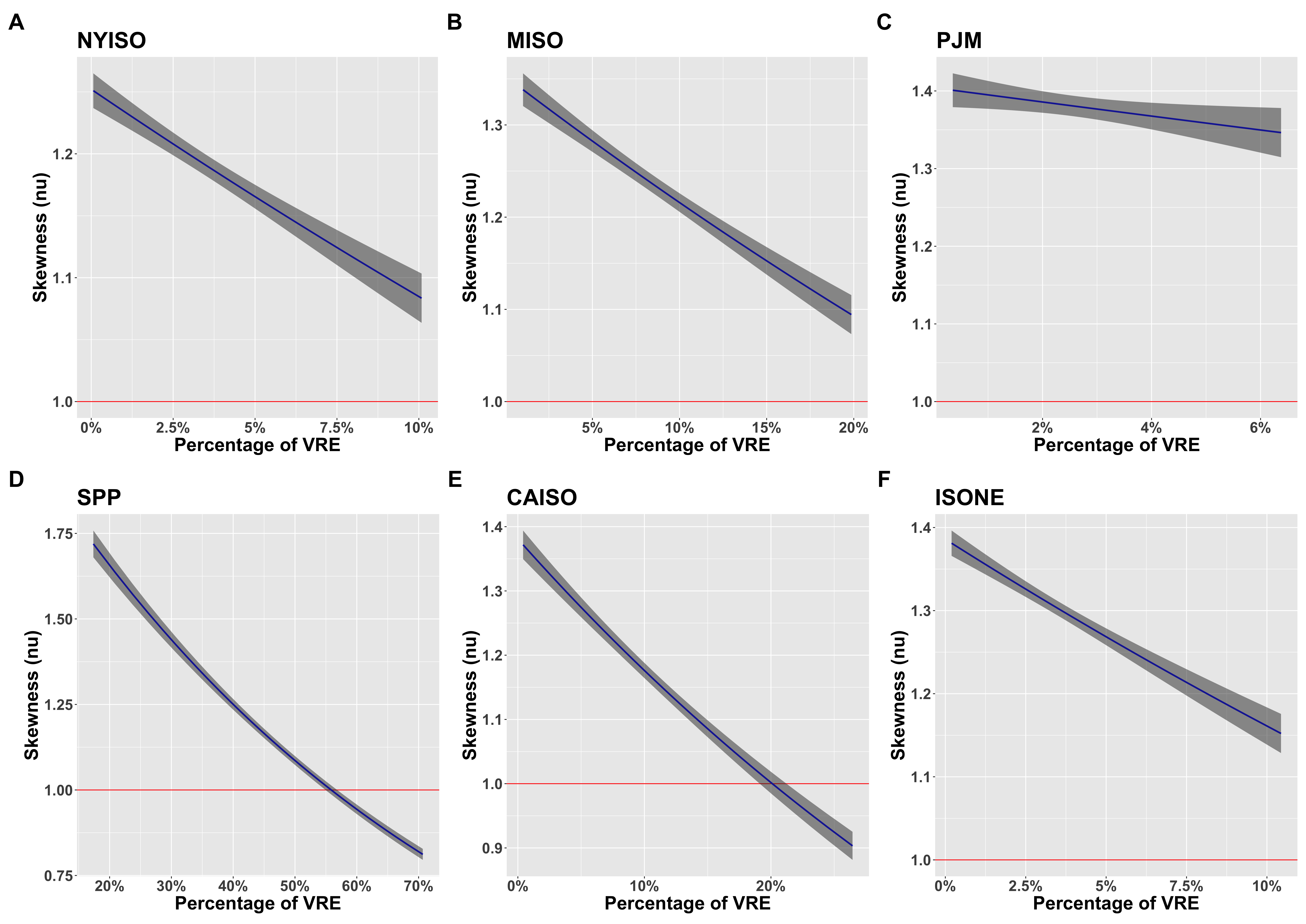}
\caption{Estimated effects of VRE percentage on the skewness ($\nu$) parameter for a skew t-distribution  for detrended price using a log link. Gray bands indicate the 95\% confidence intervals. The red reference line at 1 denotes symmetry.}
\label{fig:sst_coef}
\end{figure}

The results from the skew-t distribution regression in Figure \ref{fig:sst_coef} also show the distributional effects of increasing VRE on price.  A decreasing trend in skewness with VRE is estimated for all ISOs. The skewness approaches symmetry \added{($\nu = 1$)} with increasing VRE for four of the ISOs suggesting that as VRE is increased there is a lower frequency of large price extremes above the average price. \added{Large price extremes often happen when expensive energy generators need to be dispatched, typically when low-cost electricity generators, such as solar and wind, are unavailable or unable to meet higher than expected demand.} The magnitude and duration of these price spikes are particularly harmful to electricity retailers who cannot pass on price risk to customers \citep{anderson2007forward}. The main contribution to the pattern towards symmetry is the decrease in large, positive fluctuations from the average price. A decrease in skewness to negative is found for SPP and CAISO which implies that as VRE increases in these systems it is more likely to have deviations of \deleted{detrended} price below average. It is worth noting that these two regions (SPP and CAISO) also have the highest levels of VRE penetration. 

\added{One policy mechanism that has been used to reduce price spikes is to implement price caps. Details of the price cap including the capped value, when the price cap is enforced, and which technologies are restricted by the cap varies with time and across ISOs. Temporary price caps have been implemented in times of extreme weather events or major distortion in electricity markets} \citep{sirin2022price}. \added{In 2016, the Federal Energy Regulatory Commission issued order 831, which generally created a price cap of \$1000/MWh to be applied across all regional transmission organizations and ISOs} \citep{FERCorder}. \added{Please note that there are exceptions indicated in the order that reflect a resource’s verified cost-based incremental energy offer. To assess the impact of price caps on our analysis, we calculated the fraction of electricity prices at or above \$1000/MWh for each ISO. We found that for all ISOs, the relative frequencies were all around or less than 0.01\%. Since the price cap has been set well above the marginal and opportunity costs of all or nearly all resources, this policy mechanism has rarely prevented voluntary market clearing that would otherwise occur} \citep{WILSON200033}. \added{Therefore, the results found in our study are unlikely to be a result of price caps.}

\replaced{
The results for the system price reduction are consistent with the merit order effect of renewables where conventional generation assets are unprofitable in the market when there is a high penetration of VRE. This occurs because VREs tend to have lower marginal operating costs and are, therefore, prioritized in satisfying demand.}{The results for the system price reduction are consistent with the merit order effect. While there are regional differences in the market bidding structure, dispatch generally is guided by the merit order effect. The merit order is a way of ranking electricity generation by ascending bid price, prioritizing lower cost electricity for dispatch. The intersection of the electricity demand curve with the merit order is the clearing price, which is the price that all generators selected based on merit order will receive to produce electricity. Conventional generation assets are unprofitable in the market when there is a high penetration of VRE. This occurs because VREs tend to have lower marginal operating costs and are, therefore, prioritized in satisfying demand and selected for dispatch. As more VREs are participating in the market, there is typically a decrease in clearing prices as a result of the merit order effect.} This effect has been well studied and observed in electricity markets in several countries \citep{Sensfub2008,Cludius2014,Wurzburg2013,Cutler2011,Bockers2013green,Forrest2013,Amor2014,Huisman2013,Swinand2015,KlingeJacobsen2010,Shcherbakova2014,Winkler2016}. The lowering of electricity prices as VRE penetration increases is advantageous for consumers but not necessarily for the generators/suppliers since this causes a reduction in the income from electricity sold in the market. However, the lower cost electricity can attract energy-intensive industries, increasing electricity demand within the ISO, allowing generators/suppliers to increase revenues from increased production. Additionally, not all generators are affected equally and some, particularly those with flexible generation assets, may benefit from increased VRE penetration since they can contribute flexible reserves to manage VRE variability\citep{InternationalRenewableEnergyAgency2019,Akrami2019}. Apart from attaining a sustainable price level \citep{Winkler2016} which is good for end users, the merit order effect can also help to promote flexibility in the market to incorporate flexible generation assets that can ramp up and down in response to increasing VRE penetration. It can also incentivize storage and demand response technologies. Security of supply is another concern since the merit order effect may lead to the shut down of \added{baseload} conventional plants; however, this risk can be reduced with the introduction of capacity markets \citep{Winkler2016}. Lowering the system electricity price may increase the cost of some sustainable energy policies, such as feed-in-tariffs \citep{Huisman2013}, and may lead to lower cost flexible natural gas plants replacing nuclear plants and, consequently, producing more emissions \citep{Rothwell2000}. Careful policy design will be needed to mitigate these effects.

\begin{figure}[h!]
\centering
\includegraphics[width=14cm]{ 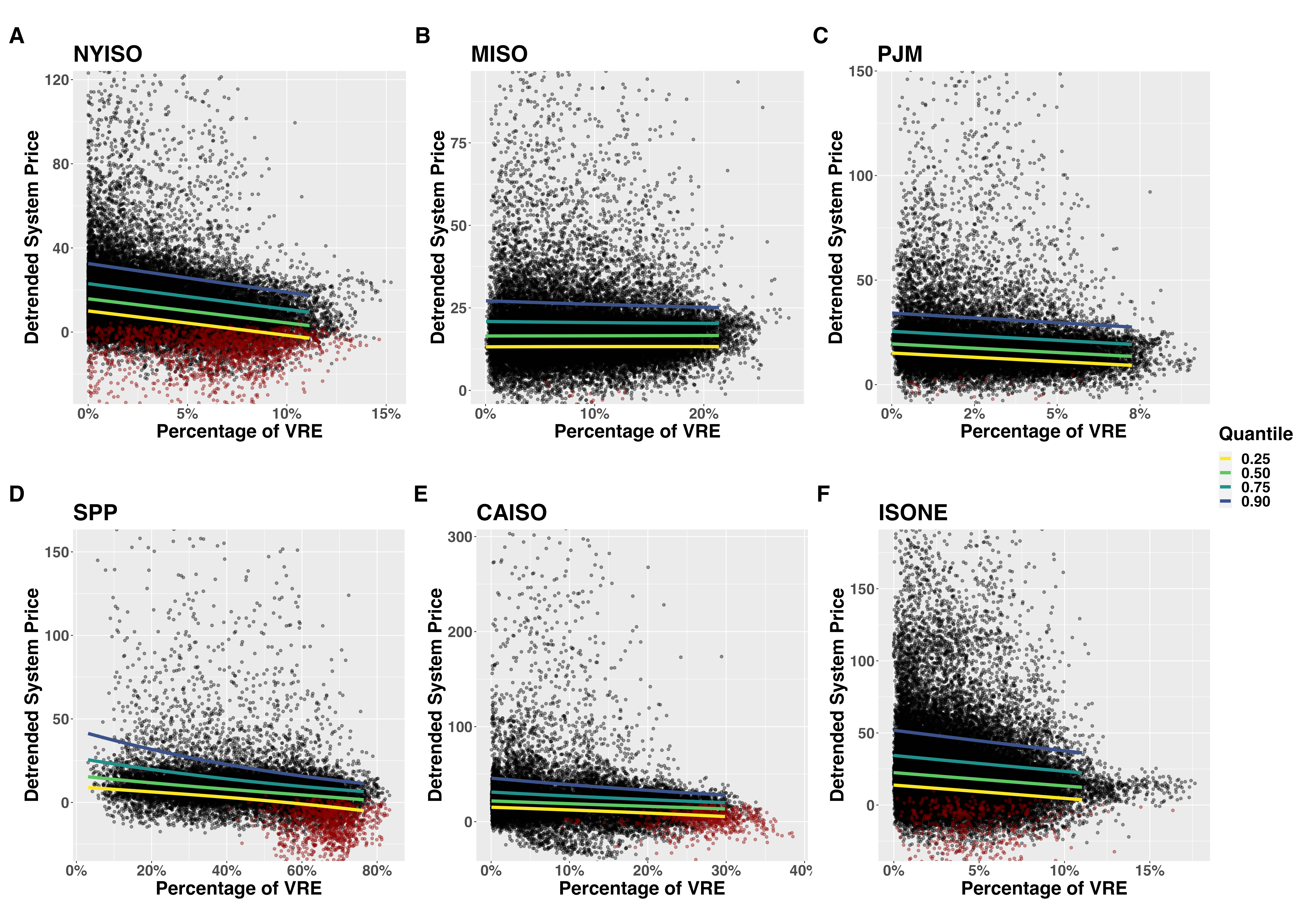} % need to fix: points need to be black/red
\caption{Quantiles of skew t-distribution based on the estimated pointwise means of the parameters listed in Table A1. Points correspond to seasonally detrended system price and y axis is trimmed to inner 99\% of detrended price; red points correspond to a negative price before seasonal detrending.} % NOTE: intercept added back and positive/negative shading added
\label{fig:sst_quant}
\end{figure}

Negative prices are the extreme end-products of oversupply of electricity. It is most common in markets with large amounts of nuclear, hydroelectric, and wind generation\citep{EIA2012}. 
However, conventional resources also play a role in driving down prices (at least in the short term) as generators from these inflexible assets often prefer not to shut down or reduce their output when there is high amounts of VRE because of technical and economic costs related to output reduction and regulation costs \citep[in the case of hydroelectric power - for compliance with environmental regulations for water flow to maintain fish population;][]{EIA2012}. The trend of negative system price (based on the original raw data without correction for temporal effects) with increasing VRE penetration varies for different ISOs (shown in the red colored points in Figure \ref{fig:quantile_price} and Figure \ref{fig:sst_quant}). However, the frequency of negative detrended system price \replaced{has no clear trend with percentage of VRE; although, it seems to be concentrated at high levels of VRE penetration for SPP and CAISO, the two ISOs with the highest range of VRE penetration.}{decreases with VRE penetration for all ISOs studied except SPP.} 

This observation might suggest that at the current levels of VRE penetration, among other factors, increased VRE penetration is not solely responsible for increased frequency of negative system electricity prices but rather due to a combination of other factors such as seasonal, daily and weekend temporal effects as well as geographic, demographic, climatic and behavioral factors which are inherently different across each of the ISOs studied. In general, at very high penetration of VRE, there needs to be increased flexibility of the electrical grid in terms of availability of grid-scale energy storage systems that can act as control reserves in times of excess supply of VRE.

\section{VRE Penetration and System Electricity Price Volatility}

The system electricity price varies as a function of both time and VRE penetration. Firstly, to calculate the volatility of the detrended system electricity price as a function of time (temporal price volatility), the Exponential Weighted Moving Average (EWMSD) volatility is employed. \deleted{- an approach that uses a penalized model with weights that decrease in time and compare the temporal price volatility with the VRE penetration.
The temporal price volatility measure used in this study has roots in historical volatility popularly used in finance.  This method of calculating volatility is based on the operation stage volatility (realised effects) and should not be confused with planning stage volatility that attempts to capture differences between the predicted and observed real-time price variations }  Secondly, to assess the variability of the detrended system electricity price as a function of VRE penetration, the distributional characteristics of the skew-t model are evaluated.  

\begin{figure}[h!]
\centering
\includegraphics[width=14cm]{ 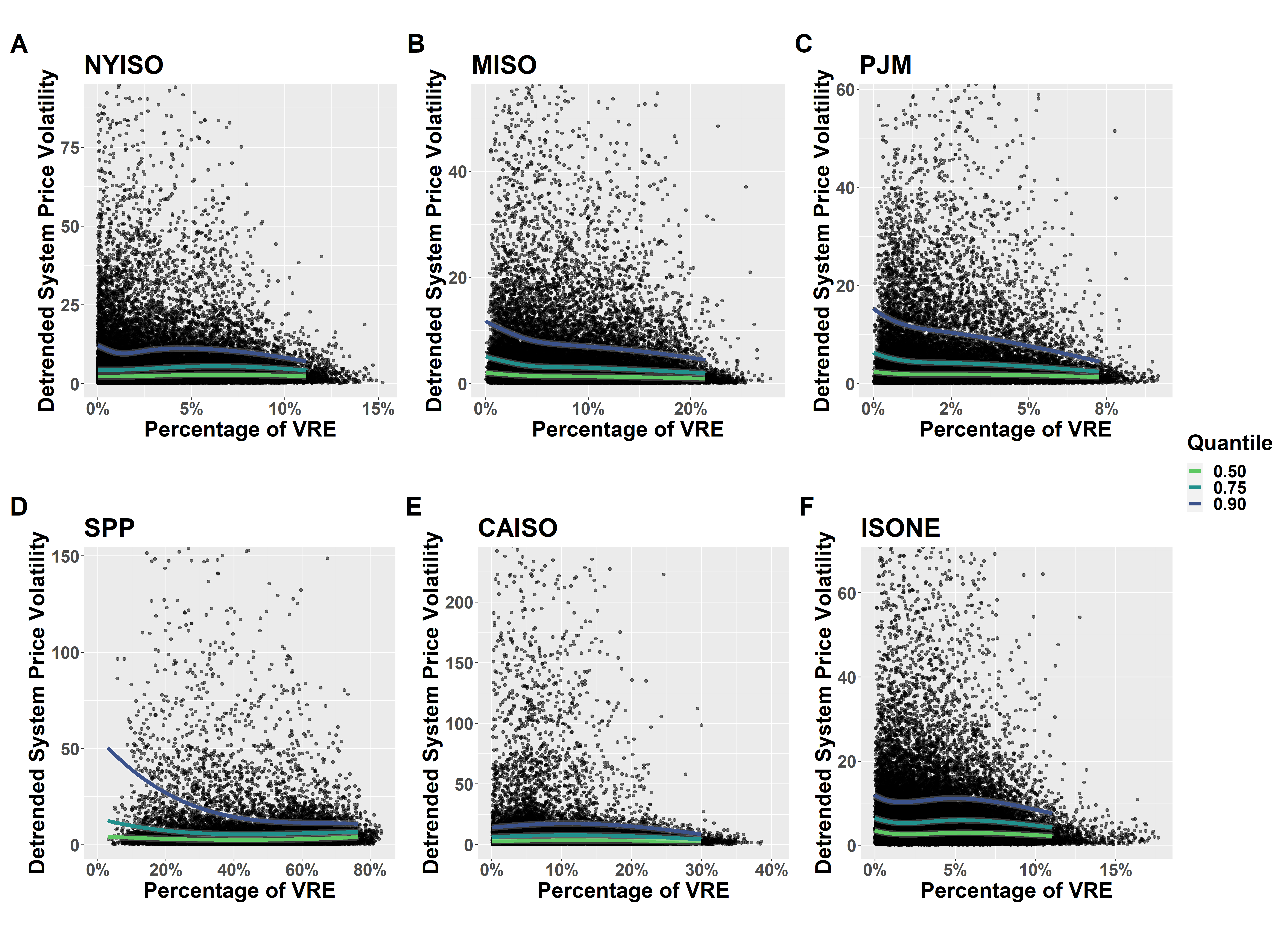}
\caption{VRE penetration (in \%) and temporal price volatility (in \$). A, B, C, D, E and F show the relationship between VRE penetration and temporal detrended system price volatility for NYISO, MISO, PJM, CAISO, ISONE and SPP respectively. The continuous lines show the quantiles with the colors - yellow, green and purple representing the \replaced{0.50th (median), 0.75th and the 0.90th quantiles}{50th (median), 75th and the 90th percentiles} respectively.}
\label{fig:quantile_volatility}
\end{figure}

The results for the quantile regression on EWMSD price volatility (temporal volatility), shown in Figure \ref{fig:quantile_volatility}, indicate that there is a reduction in temporal price volatility as the penetration of VRE increases across all quantiles \replaced{(0.50th, 0.75th and 0.90th)}{(50th, 75th and 90th)} for 5 out of the 6 ISOs studied (ISONE, MISO, PJM, CAISO, and NYISO). In the upper quantile (\replaced{0.90th quantile}{90th percentile}), it is evident that at higher quantiles there is a larger reduction in temporal price volatility as VRE penetration increases indicating that there is a significant drop in higher extremes of temporal price volatility with VRE penetration. For SPP, an increase in VRE percentage \deleted{is observed} up to about 40\% \replaced{is associated with decreasing}{decreases} the temporal price volatility; however, an increase in temporal price volatility is observed as VRE penetration increases beyond 40\%\deleted{up to about 60\%}. The difference in the effect of VRE penetration on temporal price volatility for SPP can be attributed to times of high penetration of wind when demand is low causing oversupply of electricity and, thus, increasing the frequency of negative prices. This effect observed in SPP corroborates some of the literature on price volatility where high penetration of wind can cause an increase in price volatility \citep{Rintamaki2017}. 

The change in the system temporal price volatility is not constant across the percentage of VRE nor the quantiles and, once again, this is typically more apparent at higher quantiles. For example, in PJM when the VRE percentage changes from 0-1\% the median system temporal price volatility decreases by \$0.8 USD in average absolute values and the \replaced{0.90th quantile}{90th percentile} of the system temporal price volatility  decrease by \$5 in average absolute values but when the VRE percentage changes from 5-6\% the median system temporal price volatility decreases by about \$0.2 in average absolute values and the \replaced{0.90th quantile}{90\%} temporal price volatility decreases by about \$0.4 in average absolute values. The derivative (measure of change) of the system temporal price volatility with respect to the percentage of VRE for PJM is depicted in Figure \ref{fig:derivative}.

The result of the skew-t regression elucidates the variability of detrended system electricity price \added{by estimating the energy price distribution} as a function of VRE penetration. \added{We evaluated the changes in the shape of the distribution as VRE penetration increases by investigating the differences among various quantiles. The difference between the \replaced{0.75th and 0.25th}{75th and 25th} quantile measures variability about the median detrended energy price and as it decreases, we expect less deviation from the median. The difference between the \replaced{0.75th and 0.50th}{75th and 50th} quantile and between the \replaced{0.90th and 0.50th}{90th and 50th} both show changes in above average prices where the latter describes frequency of more ``extreme" prices.} 

\begin{figure}[h!]
\centering
\includegraphics[width=14cm]{ 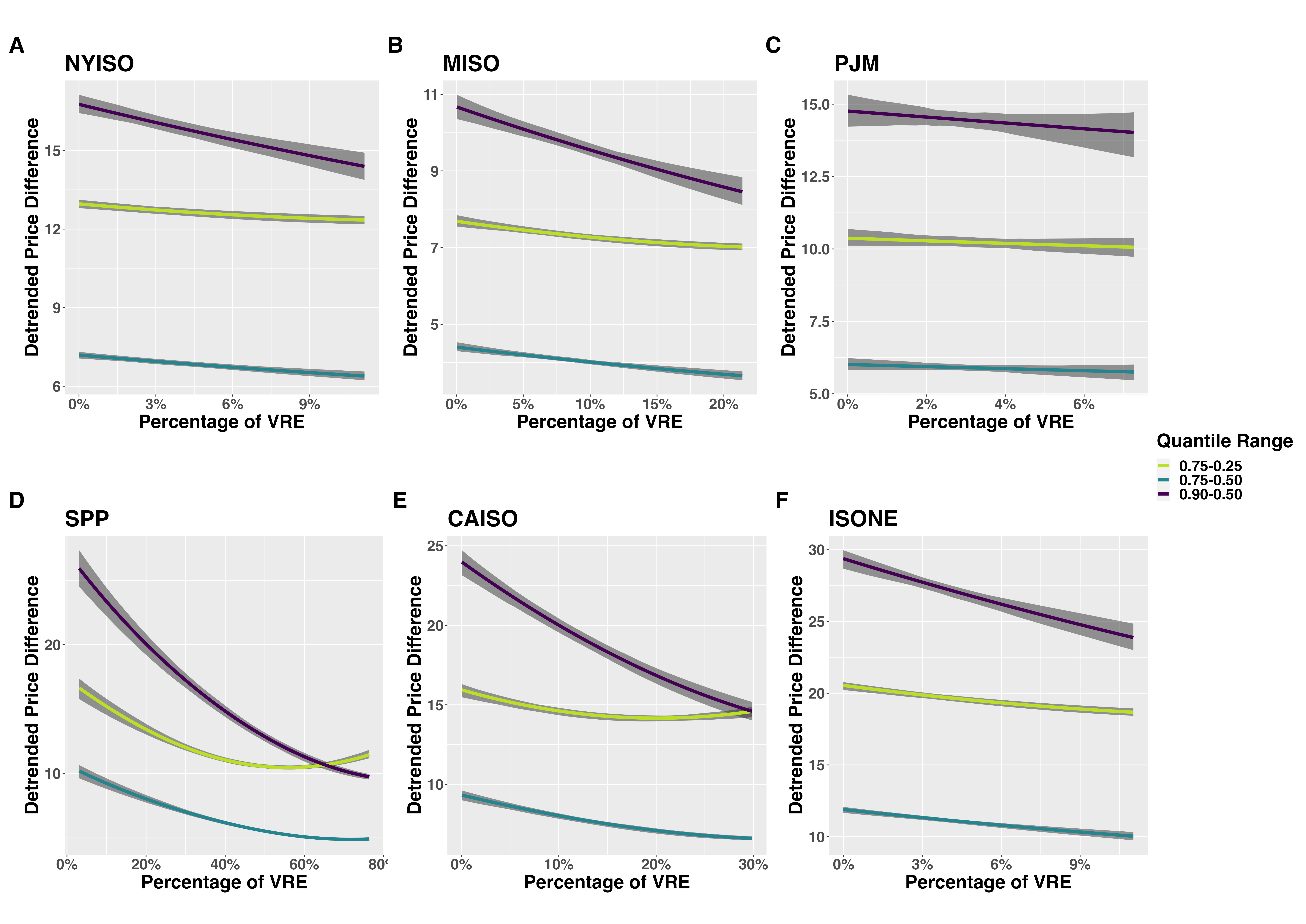}
\caption{Difference between \replaced{0.75th and 0.25th} {$75th$ and $25th$}quantile (yellow), \replaced{0.75th and 0.50th}{$75th$ and $50th$} (teal), and \replaced{0.90th and 0.50th}{$90th$ and $50th$} (purple) as plotted with approximate 95\% credible intervals. It is important note that for several of the ISOs, a decrease in the spread of the distribution (volatility with respect to VRE) is observed in the quantiles as measured by all three differences, but SPP and CAISO have curvilinear trends in the difference between \replaced{0.75th and 0.25th}{$75th$ and $25th$} which is caused by the skewness changing from positive to negative with increasing VRE penetration.}
\label{fig:sst_iqrange}
\end{figure}

In Figure \ref{fig:sst_quant}, the range of the estimated skew-t distribution quantiles for detrended system price \deleted{(a measure of price volatility relative to VRE)} decreases as percentage of VRE increases. The tightening of the distribution around the median \added{detrended} price as the percentage of VRE increases is a result of the decrease in estimated skewness parameter as VRE increases. \added{To illustrate more clearly, select} differences in the detrended system price quantiles are shown in Figure \ref{fig:sst_iqrange}. In all ISOs studied, the detrended system electricity price difference between the \replaced{0.75th and 0.50th quantile}{75th and 50th percentile} (teal) and between the \replaced{0.90th and 0.50th quantile}{90th and 50th percentile} (purple) decrease as the percentage of VRE increases. In all but two ISOs (SPP and CAISO), a reduction in the detrended system price difference between the \replaced{0.75th and 0.25th quantile}{75th and 25th percentile} (yellow) is observed as the percentage of VRE increases. From Figure \ref{fig:sst_coef}, high VRE penetration leads to negative skewness in SPP and CAISO. 
This negative skewness effect can also be seen in Figure \ref{fig:sst_quant} for the \added{0.}25th quantile (yellow) with negative deviating away from the median at high VRE penetrations in SPP and CAISO. The effect is most apparent in SPP where there is high frequency of negative prices at VRE penetrations above 40\%. 

Overall, the reductions in these multiple measures of price volatility with respect to VRE indicate that generally increased VRE penetration reduces detrended system electricity price variability, particularly through reducing extremely high electricity prices. This trend is observed in all ISOs studied for detrended system prices above the median. The exception to this trend occurs in SPP and CAISO at high penetration of VRE for detrended system prices below the median, where there is an increased frequency of negative system prices. 

\section{Conclusion}
In this study the relationship between detrended system electricity price, price volatility, and VRE penetration was analyzed using a robust approach to account for non-linearities and skewness. While several studies have investigated the relationship between VRE and electricity price using multivariate linear regressions, these are insufficient at adequately capturing the underlying relationships in the highly skewed data. The results not only corroborate the existing literature on the merit order effect of VRE, which causes a reduction in electricity price with increased VRE penetration, but also illustrates that the merit order effect is nonlinear and the greatest effect can be seen in the reductions in extremely high detrended system electricity prices (\replaced{0.90th quantile}{90th percentile}) with increased VRE. In all ISOs studied, the spread in extreme high system prices, as measured by the price difference between the \replaced{0.90th and 0.50th quantiles}{90th and 50th percentiles} and the \replaced{0.75th and 0.50th quantiles}{75th and 50th percentiles}, reduces as the percentage of VRE increases. In all but one ISO studied, system temporal price volatility (calculated based on the time-dependent EWMA price volatility) decreased as the penetration of VRE increased across all quantiles (\replaced{0.50th, 0.75th and 0.90th}{50th, 75th and 90th}).

While this study has examined the relationship between VRE penetration and the detrended system electricity price and volatility using robust techniques, it is important to comment on the bounds of validity of the study findings. The results presented in this work is valid within the studied temporal range (shown in Table \ref{tab:datasummary}) and VRE penetration levels (shown in Figure \ref{fig:VRE}). Given global and regional ambitious renewable energy targets, future VRE penetration are likely to change \deleted{in the future} to match these targets and therefore the relationship between VRE penetration may differ from the levels observed in the study result. Also, there are geographical, policy and demographic factors at play within each of the ISO studied. Direct application of the study findings should consider these interacting factors. However \deleted{despite this}, the current study provides a robust holistic understanding of the relationship between VRE penetration and the system electricity price and volatility at a national scope (6 out of 7 US ISOs) and  based on historical data which can be adapted to future analyses of this relationship and potentially serve to inform electricity market policy and decision making.

The results are consistent with the modern portfolio theory that shows that adding diverse and uncorrelated (or low correlated) assets to a portfolio is associated with a  reduction of total price volatility.  This is a particularly preferable outcome when individual assets are highly risky (i.e., individual volatilities are large), resulting in more stable and less risky portfolios of such assets. ``Diversity" is used here as a way to connect the study finding with a theoretical framework in finance where diversity is synonymous to the increase in penetration of VRE. Also, it is important to also note that the bounds of validity in relationship to diversity are based on the current VRE penetration levels and should not be extrapolated to future data. Even at that, we find that some ISOs (SPP and CAISO) show an increasing volatility at high penetration levels of VRE (relative to other regions) depending on the observed quantiles. This underscores the importance of grid-scale energy storage systems which may help to seamlessly integrate high levels of VRE thus improving grid flexibility while allowing for reserves that can balance supply in times of high availability of stochastically variable VRE resources. 

With increasing generation from VRE on the electrical grid, the use of a robust approach helps to expose the non-linear relationships %and regime switching properties 
that are useful in developing an accurate understanding of behavior of system electricity price and its volatility. However, it is important to recognize that the power grid is highly dynamic and, therefore, this study should not be construed as an argument for reaching a threshold VRE penetration. The technologies, policies, and markets associated with the grid are rapidly changing. Most regions have ambitious VRE targets that will need adequate energy storage capabilities to reduce energy curtailment and negative electricity prices, innovative markets to appropriately value demand response and auxiliary services, and mindful policies to ensure that long-term infrastructure build-out meets global sustainability goals.  
\bibliographystyle{agsm}

\section*{Acknowledgments}
\textbf{Funding:} \\
NSF Harnessing the Data Revolution (HDR) program, ``Collaborative Research: Predictive Risk Investigation SysteM (PRISM) for Multi-layer Dynamic Interconnection Analysis" (\#1940176, 1940223, 1940276, 2023755). U.S. DOE, ``Wave Energy Technology Assessment for Optimal Grid Integration and Blue Economy Advancement" (DE-EE9443).\\
\textbf{Author Contributions:}\\
Conceptualization: MGS, DAS\\
Methodology: SER, OOO, TLJS, MGS, SER, GES, SS, LW, DSM, DAS\\
Investigation: OOO\\
Visualization: OOO, TLJS, SER, GES, SS\\
Data Curation: OOO, SER\\
Formal Analysis: OOO, TLJS, MGS, SER, GES, SS, LW, DAS\\
Funding acquisition: MGS, LW, DSM, DAS\\
Project administration: OOO, MGS, DAS\\
Supervision: TLJS, MGS, LW, DSM, DAS\\
Writing – original draft: OOO, TLJS, MGS, SER, SS, LW, DAS\\
Writing – review \& editing: OOO, TLJS, MGS, GES, SS, LW, DSM, DAS\\
\textbf{Competing Interests:}\\
Authors declare no competing interest\\
\textbf{Data and materials availability:}\\
Raw data and code can be found here: \textit{https://github.com/Qunlexie/VRE-Impact-on-Price-and-Volatility}

\clearpage
\appendix
\section{Appendix}

\subsection{ISO Specific Procedures}\label{sec:iso}

From each ISO, we collected historical data on the following information:
%The data collection process involved collection of historical data that contains two main pieces of information:
\begin{enumerate}
\item Electricity Prices: This includes Locational Marginal Price, Congestion Price and Losses
\item Generation Mix Data: This includes electricity generation, separated by energy source, used to satisfy demand.%is a break down of energy sources used to satisfy demand 
\end{enumerate}
This data collection process was different for each of the ISOs since the ISOs differ in their methods of reporting market data on energy prices and operational data on their generation mix. A step-by-step of the data collection procedure for each of the ISOs is as follows:
\begin{enumerate}
\item  ISONE: The ISONE data was obtained from the ISONE API web services API v.1.1 \citet{ISONE2021} that gives a range of public market and energy data. The generation data was obtained for the entire system in resolutions ranging from 5-15 minutes and queried using [\textit{/genfuelmix}] and was then aggregated to hourly resolution. The locational marginal price data was obtained for the single hub location in hourly resolutions alongside the losses and congestion cost and queried using [\textit{/hourlylmp/rt/\\final/day/{day}/\\
location/{locationId}}]. The system price, referred to here as the marginal energy cost, is then obtained as: 
\begin{equation} \label{eq. LMP}
    MEC = LMP - MCC - MLC 
\end{equation} where:  $LMP$ = Locational Marginal Price\\
 $MEC$ = Marginal Energy Cost\\
    $MCC$  = Marginal Congestion Cost\\
    $MLC$ = Marginal Loss Cost
 
\item    NYISO: NYISO provides a repository for market report and info via their public management information system \citet{NYISO2020}. The scraper was built to obtain CSV files for the 5 minute real time system generation in NYISO and queried using [\textit{/csv/rtfuelmix/\{date\}rtfuelmix\_csv.zip}]. The locational based marginal price, congestion cost and loss cost were also obtained for each of the 11 hub locations in 5 minute resolutions and queried using 

[\textit{/csv/realtime/\{date\}realtime\_zone\_csv.zip}]. 
We found that the 4 interface locations in the NYISO have significant price variations from the internal locations and were thus excluded. We then obtain the system price, referred to here as the marginal energy cost, as:
\begin{equation}
    MEC = LBMP +MCC - MLC 
\end{equation} where:  $LBMP$ = Locational Based Marginal Price\\
% $MEC$ = Marginal Energy Cost\\
%     $MCC$  = Marginal Congestion Cost\\
%     $MLC$ = Marginal Loss Cost

It is important to note that price calculations in NYISO \citet{NewYorkISO2020} is slightly different from all the other ISOs . The 5 minute resolution data was then mean-aggregated to an hourly resolution which was used in the analysis.

\item PJM: PJM has an interactive data platform, Data Miner 2 \citet{PJM2021}, where market and operational data is stored. The platform provides hourly locational marginal price, marginal loss price and congestion price which can be queried [\textit{/feed/rt\_da\_monthly\_lmps}].
The system real time price was also provided directly but can also be obtained using (Eqn. 1) above. The generation mix data was obtained in hourly resolution for the entire PJM ISO and queried using [\textit{/feed/gen\_by\_fuel}]
\item MISO: MISO provides a market and operational report repository \citet{MISO2021} which we directly scraped. The generation mix data was obtained from this repository in hourly resolution and queried using [\textit{/marketreports/\\\{day\}\_sr\_gfm.xls}]. The locational marginal, congestion and loss prices were also obtained for each of the hub and queried using [\textit{/marketreports/\{day\}\_rt\_lmp\_final\_csv.zip}]. Similarly, a unique system price was calculated using  (Eqn. 1) above.
\item SPP: SPP provides an integrated marketplace \citet{Spp2021} in the form of a file browser API platform for downloading market reports. The scraper was then built to download the provided 5-minute resolution generation data  and queried using [\textit{/generation-mix-historical}]. The same procedure was used to obtain the 5-minute locational marginal prices, congestion price and the loss price and queried using [\textit{rtbm-lmp-by-location}] for specified hub locations after which a unique system price was obtained (using Eqn. 1). 
\item CAISO: The scraper was built to obtain data from the California ISO Open Access Same-Time Information System (OASIS) API via schema \citet{CaliforniaISO2017} that use structured query parameters to fetch data from the system. The data collected includes: 5-minute real time renewables (solar and wind) forecast by trading hub,  5-minute real time total generation by entire CAISO area, and 5-minute interval local marginal price, marginal loss price and the congestion price by trading hub. Similarly, a unique system price was calculated using  (Eqn. 1) above. The renewable generation data was then sum-aggregated for all hub locations in order to obtain a representation for the total CAISO region. 

\end {enumerate}

\subsection{Detrending Example}\label{appendix:detrend}

\begin{figure}[h!]
\centering
\includegraphics[width=14cm]{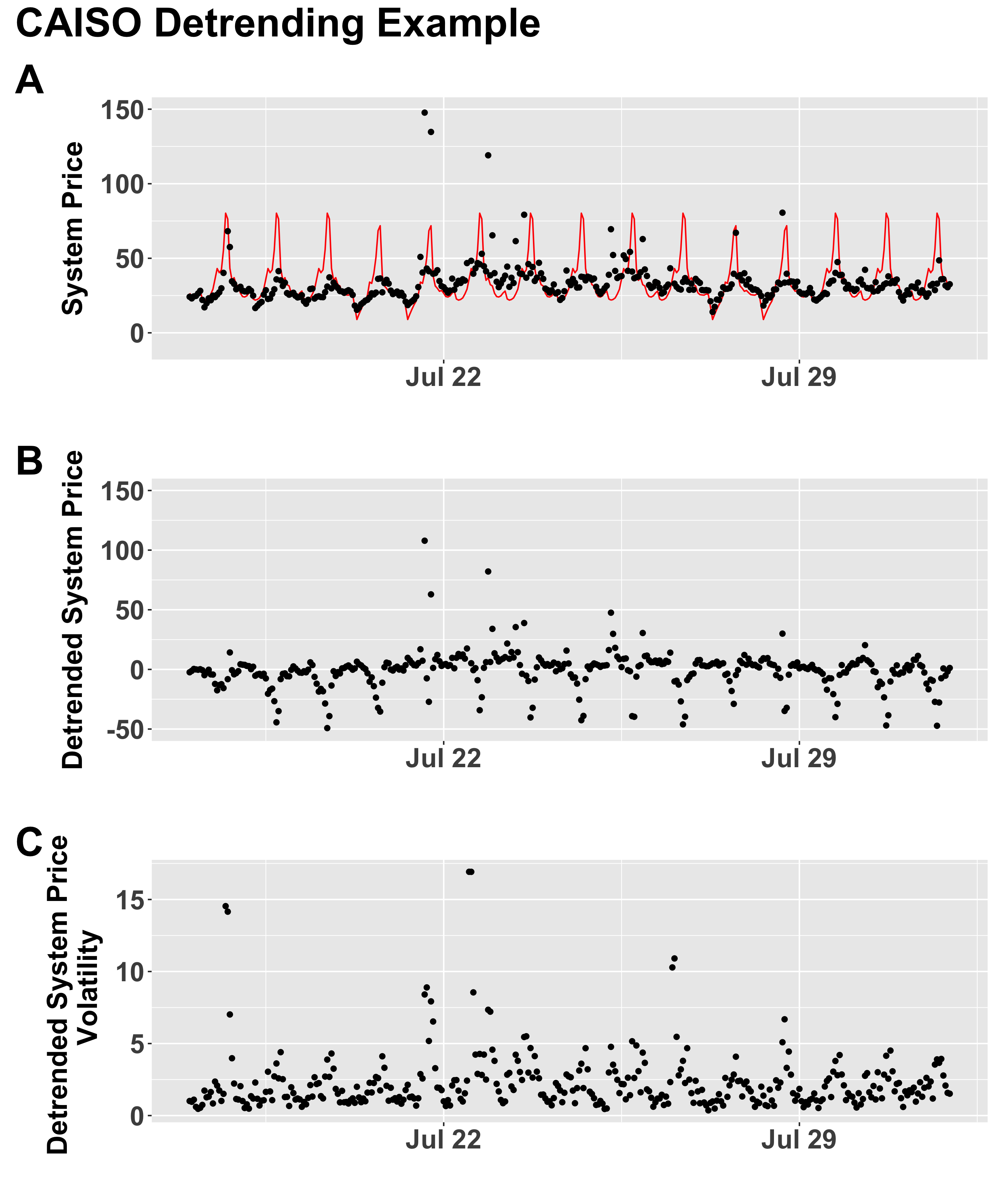}
\caption{The three panels show an illustration of the detrending procedure for a snapshot of data from CAISO in the second half of July 2019. In (A) the raw system price is shown (dots) along with the estimated seasonal trend (red line). We detrend the raw system price by subtracting the trend to obtain the detrended system price (B). The detrended system price represents the how much the system price differs from the seasonal average and notice most of the time it is near 0. From (B), we calculated the exponentially weighted moving average standard deviation (C) that is a temporal measure of the variability of the detrended system price. In other words, how much does the detrended price vary from the detrended prices in the recent past.}
\label{fig:detrend_example}
\end{figure}

\clearpage{}

\subsection{Tables}

\begin{table}[]
\renewcommand{\thetable}{A1}
    \centering

\begin{tabular}{llrr}
\toprule
ISO & parameter & Estimate & Std. Error\\
\midrule
 & $\beta_0^{\mu}$, centrality intercept & -6.68 & 0.11\\

 & $\beta_1^{\mu}$, change in centrality for a change in VRE & -7.96 & 0.87\\

 & $\beta_0^{\sigma}$, log-scale estimate & 2.14 & 0.01\\

 & $\beta_0^{\nu}$, skewness intercept & 0.32 & 0.01\\

 & $\beta_1^{\nu}$, change in skewness for a change in VRE & -1.61 & 0.07\\

\multirow{-6}{*}{\raggedright\arraybackslash CAISO} & $\beta_0^{\tau}$, log-tail index estimate & 0.59 & 0.01\\
\cmidrule{1-4}
 & $\beta_0^{\mu}$, centrality intercept & -7.46 & 0.11\\

 & $\beta_1^{\mu}$, change in centrality for a change in VRE & -61.89 & 2.54\\

 & $\beta_0^{\sigma}$, log-scale estimate & 2.42 & 0.00\\

 & $\beta_0^{\nu}$, skewness intercept & 0.33 & 0.01\\

 & $\beta_1^{\nu}$, change in skewness for a change in VRE & -1.77 & 0.14\\

\multirow{-6}{*}{\raggedright\arraybackslash ISONE} & $\beta_0^{\tau}$, log-tail index estimate & 0.72 & 0.01\\
\cmidrule{1-4}
 & $\beta_0^{\mu}$, centrality intercept & -4.04 & 0.05\\

 & $\beta_1^{\mu}$, change in centrality for a change in VRE & 7.53 & 0.53\\

 & $\beta_0^{\sigma}$, log-scale estimate & 1.47 & 0.01\\

 & $\beta_0^{\nu}$, skewness intercept & 0.30 & 0.01\\

 & $\beta_1^{\nu}$, change in skewness for a change in VRE & -1.07 & 0.08\\

\multirow{-6}{*}{\raggedright\arraybackslash MISO} & $\beta_0^{\tau}$, log-tail index estimate & 0.78 & 0.01\\
\cmidrule{1-4}
 & $\beta_0^{\mu}$, centrality intercept & -0.36 & 0.07\\

 & $\beta_1^{\mu}$, change in centrality for a change in VRE & -98.49 & 1.62\\

 & $\beta_0^{\sigma}$, log-scale estimate & 2.05 & 0.01\\

 & $\beta_0^{\nu}$, skewness intercept & 0.22 & 0.01\\

 & $\beta_1^{\nu}$, change in skewness for a change in VRE & -1.43 & 0.13\\

\multirow{-6}{*}{\raggedright\arraybackslash NYISO} & $\beta_0^{\tau}$, log-tail index estimate & 0.89 & 0.01\\
\cmidrule{1-4}
 & $\beta_0^{\mu}$, centrality intercept & -4.26 & 0.08\\

 & $\beta_1^{\mu}$, change in centrality for a change in VRE & -76.27 & 2.47\\

 & $\beta_0^{\sigma}$, log-scale estimate & 1.74 & 0.01\\

 & $\beta_0^{\nu}$, skewness intercept & 0.34 & 0.01\\

 & $\beta_1^{\nu}$, change in skewness for a change in VRE & -0.66 & 0.28\\

\multirow{-6}{*}{\raggedright\arraybackslash PJM} & $\beta_0^{\tau}$, log-tail index estimate & 0.75 & 0.01\\
\cmidrule{1-4}
 & $\beta_0^{\mu}$, centrality intercept & -2.31 & 0.18\\

 & $\beta_1^{\mu}$, change in centrality for a change in VRE & -4.11 & 0.36\\

 & $\beta_0^{\sigma}$, log-scale estimate & 1.87 & 0.01\\

 & $\beta_0^{\nu}$, skewness intercept & 0.79 & 0.02\\

 & $\beta_1^{\nu}$, change in skewness for a change in VRE & -1.41 & 0.03\\

\multirow{-6}{*}{\raggedright\arraybackslash SPP} & $\beta_0^{\tau}$, log-tail index estimate & 0.76 & 0.01\\
\bottomrule
\end{tabular}

\caption{Parameter values for the generalized skew t-distribution hourly detrended price. An example of interpretation of the effect of VRE on the centrality and skewness of the distribution of hourly price is an increase in VRE decreases both centrality and skewness of hourly price in CAISO  ($\beta_1^{\mu} < 0$ and $\beta_1^{\nu} < 0$). The interpretation also holds for ISONE, NYISO and SPP.}
\label{tab:skew_hour}
\end{table}

\clearpage{}
%\printbibliography
\end{document}